\documentclass{emulateapj}
\def\umu{{mag arcsec$^{-2}$}}

\shorttitle{pCMDs of Bright Galaxies in A1139 and A2589}

\shortauthors{Lee et al.}
\def\simlt{\lower.5ex\hbox{$\; \buildrel < \over \sim \;$}}
\def\simgt{\lower.5ex\hbox{$\; \buildrel > \over \sim \;$}}

\begin{document}

\title{ Color Dispersion as an Indicator of Stellar Population Complexity: Insights from the Pixel Color-Magnitude Diagrams of 32 Bright Galaxies in Abell~1139 and Abell~2589}

\author{Joon Hyeop Lee$^{1,2}$, Mina Pak$^{1,2}$, Hye-Ran Lee$^{1,2}$, Sree Oh$^{3,4,5}$}
\affil{$^1$ Korea Astronomy and Space Science Institute, Daejeon 34055, Korea\\
$^2$ University of Science and Technology, Daejeon 34113, Korea\\
$^3$ Research School of Astronomy and Astrophysics, The Australian National University, Canberra, ACT 2611, Australia\\
$^4$ ARC Centre of Excellence for All Sky Astrophysics in 3 Dimensions (ASTRO 3D), Australia\\
$^5$ Department of Astronomy and Yonsei University Observatory, Yonsei University, Seoul 03722, Korea}

\email{jhl@kasi.re.kr}

\begin{abstract}
We investigate the properties of bright galaxies with various morphological types in Abell~1139 and Abell~2589, using the pixel color-magnitude diagram (pCMD) analysis technique. The sample is 32 galaxies brighter than $M_r=-21.3$ mag with spectroscopic redshifts, which are deeply imaged in the $g$ and $r$ bands using the MegaCam mounted on the Canada-France-Hawaii Telescope. After masking contaminants with two-step procedures, we examine how the detailed properties in pCMDs depend on galaxy morphology and infrared color.
The mean $g-r$ color as a function of surface brightness ($\mu_r$) in the pCMD of a galaxy shows fine performance in distinguishing between early- and late-type galaxies, but it is not perfect because of the similarity between elliptical galaxies and bulge-dominated spiral galaxies. On the other hand, the $g-r$ color dispersion as a function of $\mu_r$ works better. We find that the best set of parameters for galaxy classification is the combination of the minimum color dispersion at $\mu_r\le21.2$ {\umu} and the maximum color dispersion at $20.0\le\mu_r\le21.0$ {\umu}: the latter reflects the complexity of stellar populations at the disk component in a typical spiral galaxy.
Finally, the color dispersion measurements of an elliptical galaxy appear to be correlated with the WISE infrared color ($[4.6] - [12]$). This indicates that the complexity of stellar populations in an elliptical galaxy is related with its recent star formation activities. From this observational evidence, we infer that gas-rich minor mergers or gas interactions may have usually happened during the recent growth of massive elliptical galaxies.
\end{abstract}

\keywords{galaxies: clusters: individual (Abell 1139, Abell 2589) --- galaxies: elliptical and lenticular, cD --- galaxies: spiral --- galaxies: evolution --- galaxies: formation}

\section{INTRODUCTION}\label{intro}

The comprehensive understanding of the formation histories of galaxies in variety of morphology is a major goal in modern astronomy. Over the classical paradigms about the formation of massive galaxies such as the monolithic collapse model \citep{pat67,tin72,lar74} and the hierarchical merging model \citep{too77,sea78}, a huge number of studies in recent decades have illuminated various aspects in the mass assembly and star formation histories of galaxies. Currently, much observational evidence supports a galaxy formation scenario that is more complicated than classically assumed: massive galaxies may have built through two-phase and inside-out formation processes \citep[e.g.,][]{van10,ose10,lee13a,pas15,liu16}. In this scenario, massive galaxies have experienced violent starburst and subsequent star formation quenching at a relatively early epoch, which partially resembles the classical monolithic collapse. On the other hand, such a violent starburst is thought to be often triggered by gas-rich major mergers as well as those galaxies are found to have kept growing via numerous minor mergers for a long time after the early starburst and quenching, which is consistent with the hierarchical build-up scheme.

Although recent observational and theoretical achievements shed light on our understanding of galaxy formation by suggesting a reasonable solution to the apparent contradiction between the classical models, however, several detailed issues are still under a veil. For example, it has not been perfectly comprehended how the recent histories of star formation and mass assembly in massive galaxies are related with each other. Several recent studies argued that the major formation process of early-type galaxies depends on their masses, in the context that low-mass early-type galaxies tend to have experienced gas-rich mergers, whereas massive early-type galaxies have grown via dry mergers \citep{kor09,ber11,lee13b}. However, whereas those findings well explain the average trends of galaxy formation, we witness various exceptions such as massive galaxies with evidence of recent star formation activities \citep[e.g.,][]{lee06,fer11,geo17,she17}.

It is relatively easy to trace current or recent star formation in a galaxy by using various well-defined indicators, such as optical or UV $-$ optical colors, spectral emission lines, and infrared or radio luminosity. On the other hand, the method to trace recent mass assembly events is not very well established, compared to star formation measurement.
One may determine whether a galaxy recently experienced merging events or not, by decomposing its internal structures and estimating its structural asymmetry or irregularity. However, such an approach requires much effort for each galaxy, as well as quantitative comparison between different galaxies is not easy.
Thus, if a simple and efficient indicator of recent mass assembly is devised, it will be very useful for investigating recent evolution histories of a large number of galaxies, in combination with the tracers of recent star formation activities.

A candidate for such a simple and efficient methodology to trace the formation history of a galaxy is the pixel color-magnitude diagram (pCMD) analysis \citep{bot86,abr99,deg03,lan07,lee11,lee12,lee17}.
This method has not been very widely used until now, but it has potential to be a powerful approach to total understanding of the photometric and structural properties of galaxies.
For example, \citet{lan07} showed that galaxies in variety of morphological types show various and distinct features in their pCMDs, such as \emph{prime sequences} for early-type galaxies and \emph{inverse-L} features for spiral galaxies.
Compared to the classical photometric and/or structural analysis methods, the pCMD analysis has several advantages: (1) it can be consistently applied to galaxies with any kind of morphology, (2) it considers the photometric and structural properties at the same time, and (3) it is efficient in checking the homology between galaxies in different mass or size scales \citep{lee17}.

Despite such merits, however, the usage of the pCMD analysis is not perfectly established yet.
As an effort to devise a quantitative analysis method using pCMDs, \citet{lee17} introduced simple routines to compare pCMDs to check the similarity in galaxy formation history. They compared the brightest cluster galaxies (BCGs) in Abell~1139 (A1139) and Abell~2589 (A2589) using the pCMD analysis technique, which are two of the 14 galaxy clusters targeted in the KASI-Yonsei Deep Imaging Survey of Clusters \citep[KYDISC;][]{oh18}. By simplifying the overall pCMD features into the mean and standard deviation of pixel color as a function of pixel surface brightness (called a pCMD \emph{backbone}), \citet{lee17} showed that the BCG of dynamically relaxed A2589 formed a larger central body at an early epoch and has grown to be a larger, more massive and dynamically better relaxed galaxy today than the BCG of dynamically young A1139. Those results support the idea that a BCG and its host cluster coevolve. 

As shown in \citet{lee17}, the pCMD backbone analysis is a useful method for comparing the formation histories of galaxies even with different mass or size scales, but it needs to be more simplified to be efficiently applied to a large number of galaxies.
One of the remarkable quantities is the color dispersion at given $\mu_r$ in a pCMD. This quantity is thought to reflect the complexity of stellar populations. That is, a dynamically young galaxy, in which multiple stellar populations are not well mixed spatially, must have large color dispersion. Based on this inference, \citet{lee17} concluded that the BCG of A1139 is dynamically younger than the BCG of A2589.
However, since BCGs usually have simple structures, it will be worth examining the performance of this parameter for more various galaxies.

This is the second of the series papers presenting the pCMD analysis results of the KYDISC cluster galaxies.
In this paper, we present our pCMD study of 32 bright galaxies in A1139 and A2589. The goals of this paper are (1) to examine how the pCMD properties depend on galaxy morphology quantitatively, and (2) to understand how the stellar population complexity of a galaxy indicated by color dispersion in a pCMD is related with recent star formation activity represented by infrared color.
This paper is outlined as follows. Section~2 describes observations and sample selection. Section~3 presents the standard and alternative procedures for masking contaminants to yield final pCMDs. The results about the pCMD properties depending on morphology and star formation activity are shown in Section~4. The implication of our results is discussed in Section~5, and finally the conclusions are drawn in Section~6. In Appendix, additional figures showing the detailed processes of masking contaminants are presented, and the final results after the alternative masking procedure are shown, which are in good agreement with the results after the standard masking procedure. 
Throughout this paper, we adopt the cosmological parameters: $h=0.7$, $\Omega_{\Lambda}=0.7$, and $\Omega_{M}=0.3$.

\section{OBSERVATIONS AND TARGETS}\label{data}

\begin{figure*}
\centering
\includegraphics[width=0.95\textwidth]{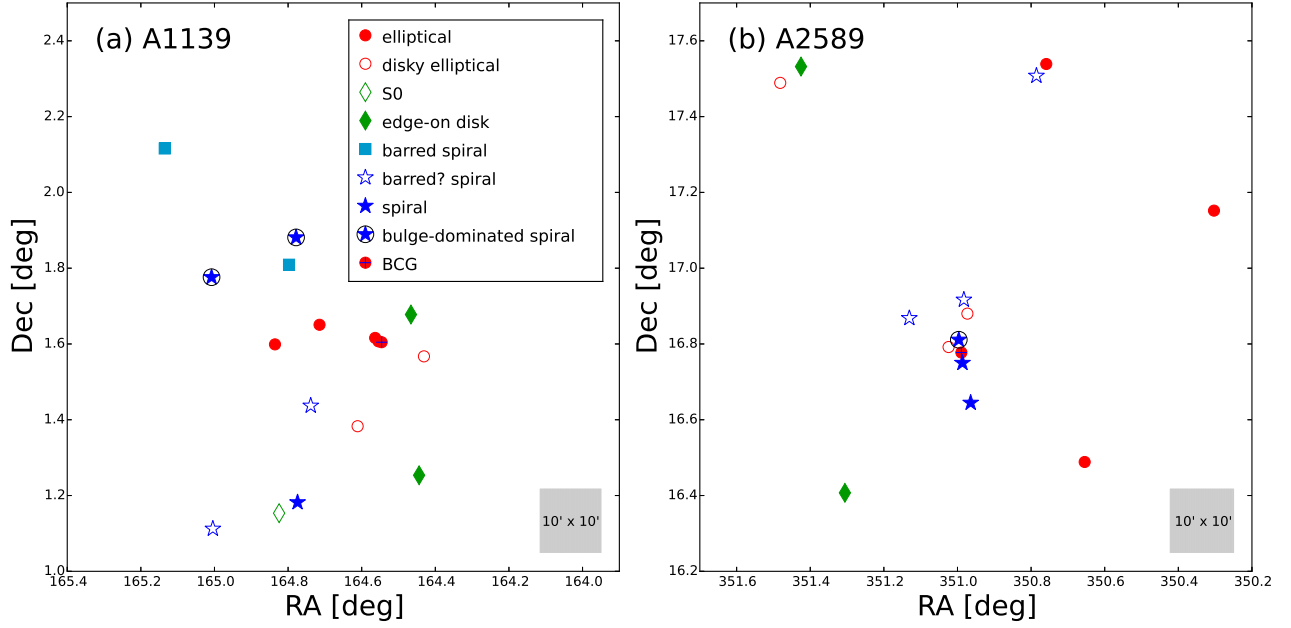}
\caption{Spatial distribution of the sample galaxies in (a) A1139 and (b) A2589. The visually classified morphological types are denoted (see Section~\ref{morcla} for the details). The gray box in each panel shows the angular scale of $10 \times 10$ square arcminutes ($\approx 0.5 \times 0.5$ Mpc$^2$).\label{spdist}}
\end{figure*}

\begin{figure*}
\centering
\includegraphics[width=0.95\textwidth]{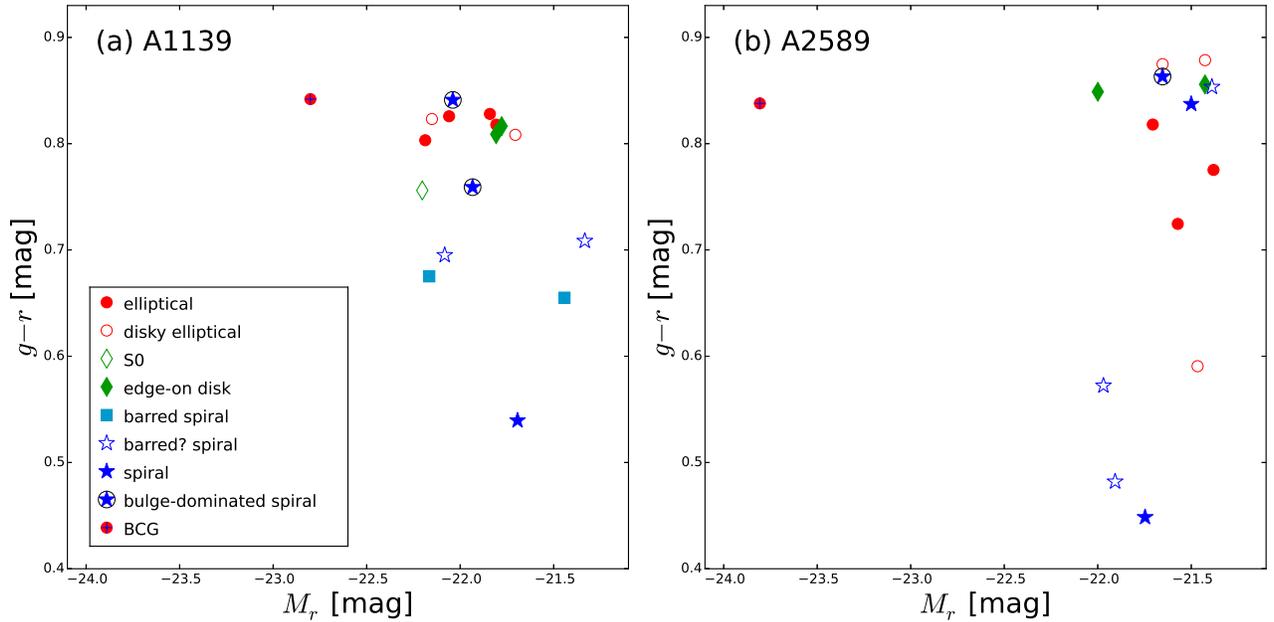}
\caption{Total color-magnitude diagrams of the sample galaxies in (a) A1139 and (b) A2589, with the morphological types denoted.\label{intcmd}}
\end{figure*}

\begin{figure*}
\centering
\includegraphics[width=0.95\textwidth]{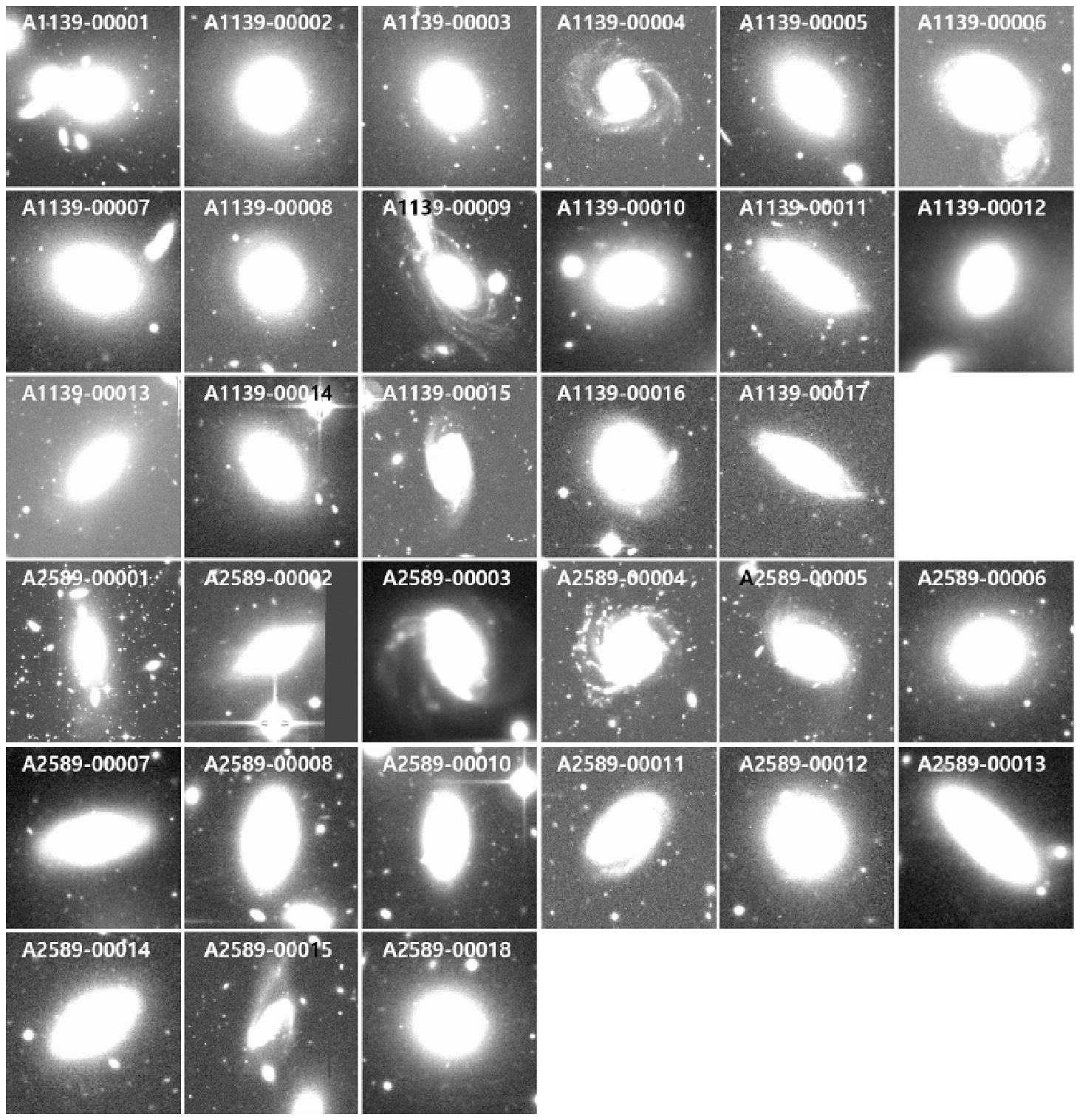}
\caption{Portrait images of the sample galaxies in the $r$ band. The name denoted at the upper side in each panel is the identity number in the KYDISC catalog.\label{portrait}}
\end{figure*}

\subsection{CFHT Observations and Sample Selection}

Our data set were obtained using the MegaCam mounted on the 3.6-meter Canada-France-Hawaii Telescope (CFHT) in 2012 -- 2013, as a part of the KYDISC project. Among the 14 nearby clusters at $0.016\le z \le 0.145$ observed in the KYDISC project, the target clusters in this paper are A1139 and A2589, which are at very similarly low redshifts ($z\sim0.04$).
This similarity in redshift makes the pCMD comparison much easier, because the galaxies in the two target clusters have almost the same conversion factors from angular size to physical scale ($\sim0.8$ kpc per arcsec) and central wavelengths at the rest frame.
See Table~1 of \citet{lee17} for the basic information of the two target clusters.

In the $g$ and $r$ bands respectively, each cluster was deeply imaged with total exposure time of 2940 seconds. After pre-processing and image stacking, the finally resampled pixel scale is $0.185''$ and the stellar full width at half maximum (FWHM) is about $0.8''$. Based on the redshift information retrieved from the Sloan Digital Sky Survey \citep{yor00}, NASA Extragalactic Database\footnote{http://ned.ipac.caltech.edu} and SIMBAD Astronomical Database\footnote{http://simbad.u-strasbg.fr/simbad/}, cluster members were selected by using the difference in recession velocity from the host cluster ($|\Delta v_{rec}| \le 3 \sigma_{cl}$) and the clustercentric distance ($R\le 2 R_{200}$), where $\sigma_{cl}$ and $R_{200}$ are the velocity dispersion and virial radius of a given cluster, respectively.
More details about the KYDISC project and its data products are described in \citet{oh18}.

In this paper, we investigate  the bright ($M_r\le-21.3$ mag) galaxies in A1139 and A2589 using the pCMD analysis method.
Among the selected cluster members, there are 16 and 15 galaxies brighter than the magnitude cut in A1139 and A2589, respectively. In addition to these 31 member galaxies, there is a bright galaxy (A1139-00004) that satisfies $|\Delta v_{rec}| \le 3 \sigma_{cl}$ but is located at $2.2 \times R_{200}$ distance from the A1139 center. This galaxy is not a member of A1139 in our selection criteria, but we include this galaxy in our sample, too. Since this paper is not focused on environmental effects, the important is the fact that this galaxy is at a distance from us similar to those of the cluster members, not its membership itself.

In this paper, we apply $k$-correction to neither pixel surface brightness nor total magnitude, because the $k$-correction based on 2-band color may not be sufficiently reliable, particularly when the photometric uncertainty is large (e.g., faint pixels).  On the other hand, \citet{lee17} did not apply $k$-correction to pixel surface brightness either, but they applied it to total magnitude when selecting cluster members for comparison with BCGs. Thus, the sample in this paper does not exactly coincide with the comparison sample in \citet{lee17}. Despite such difference from \citet{lee17}, we decided that the photometric consistency between the pixel and total magnitudes is more important in this paper.

The basic information of our sample galaxies is summarized in Table~\ref{galinfo}. The right ascensions (RAs) and declinations (Decs) are for the J2000 epoch. All magnitudes and surface brightness in this paper were corrected for the Galactic extinction using the method of \citet{sch11}, in the AB magnitude system.
Figure~\ref{spdist} presents the spatial distribution of the selected sample galaxies, while their total color-magnitude diagrams are shown in Figure~\ref{intcmd}.
The $r$-band portrait images of our sample galaxies are displayed in Figure~\ref{portrait}.

\begin{deluxetable*}{cccccccccl}
\tabletypesize{\footnotesize}
\tablenum{1} \tablecolumns{10} \tablecaption{ Basic Information of the Sample Galaxies} \tablewidth{0pt}
\tablehead{ Name & RA & Dec & Redshift & $M_r$ & $g-r$ & $|\Delta v_{rec}| / \sigma_{cl}$ & $R / R_{200}$ & B/T & Morphology \\
& [hh:mm:ss] & [dd:mm:ss] & & [mag] & [mag] & & & & }
%\rotate
\startdata
A1139-00001 & 10:58:11.0 & +01:36:17 & 0.038 & $-22.801$ & 0.842 & 0.88 & 0.27 & 0.94 & elliptical (BCG) \\
A1139-00002 & 10:59:17.9 & +01:09:12 & 0.039 & $-22.202$ & 0.756 & 0.21 & 1.15 & 0.61 & S0 \\
A1139-00003 & 10:58:51.5 & +01:39:02 & 0.040 & $-22.187$ & 0.803 & 0.04 & 0.62 & 0.82 & elliptical \\
A1139-00004 & 11:00:32.4 & +02:06:57 & 0.039 & $-22.165$ & 0.675 & 0.26 & 2.17 & 0.21 & barred spiral \\
A1139-00005 & 10:58:26.7 & +01:22:59 & 0.041 & $-22.151$ & 0.823 & 0.69 & 0.37 & 0.70 & disky elliptical \\
A1139-00006 & 10:58:57.2 & +01:26:13 & 0.041 & $-22.083$ & 0.695 & 1.12 & 0.57 & 0.26 & barred? spiral \\
A1139-00007 & 10:59:20.5 & +01:35:56 & 0.039 & $-22.059$ & 0.826 & 0.48 & 0.83 & 0.69 & elliptical \\
A1139-00008 & 11:00:01.9 & +01:46:34 & 0.040 & $-22.039$ & 0.841 & 0.30 & 1.40 & 0.42 & spiral (bulge-dominated) \\
A1139-00009 & 10:59:06.8 & +01:52:52 & 0.041 & $-21.934$ & 0.759 & 0.56 & 1.15 & 0.65 & spiral (bulge-dominated) \\
A1139-00010 & 10:58:15.2 & +01:36:58 & 0.039 & $-21.841$ & 0.828 & 0.32 & 0.31 & 0.58 & elliptical \\
A1139-00011 & 10:57:51.9 & +01:40:40 & 0.041 & $-21.808$ & 0.809 & 1.04 & 0.46 & 0.44 & edge-on disk \\
A1139-00012 & 10:58:13.0 & +01:36:25 & 0.039 & $-21.807$ & 0.818 & 0.85 & 0.28 & 0.64 & elliptical \\
A1139-00013 & 10:57:46.6 & +01:15:13 & 0.039 & $-21.778$ & 0.817 & 0.41 & 0.64 & 0.87 & edge-on disk \\
A1139-00014 & 10:57:43.4 & +01:34:02 & 0.039 & $-21.705$ & 0.808 & 0.41 & 0.28 & 0.77 & disky elliptical \\
A1139-00015 & 10:59:05.8 & +01:10:55 & 0.039 & $-21.693$ & 0.540 & 0.57 & 1.02 & 0.03 & spiral \\
A1139-00016 & 10:59:11.4 & +01:48:33 & 0.040 & $-21.442$ & 0.655 & 0.10 & 1.04 & 0.44 & barred spiral \\
A1139-00017 & 11:00:01.1 & +01:06:44 & 0.039 & $-21.334$ & 0.709 & 0.59 & 1.55 & 0.09 & barred? spiral \\
A2589-00001 & 23:23:57.5 & +16:46:38 & 0.041 & $-23.807$ & 0.838 & 0.09 & 0.04 & 0.98 & elliptical (BCG) \\
A2589-00002 & 23:25:42.0 & +17:31:55 & 0.041 & $-22.000$ & 0.849 & 0.06 & 1.04 & 0.44 & edge-on disk \\
A2589-00003 & 23:23:08.5 & +17:30:28 & 0.038 & $-21.970$ & 0.572 & 1.12 & 0.89 & 0.17 & barred? spiral \\
A2589-00004 & 23:24:31.4 & +16:52:05 & 0.036 & $-21.907$ & 0.482 & 1.89 & 0.20 & 0.39 & barred? spiral \\
A2589-00005 & 23:23:51.4 & +16:38:41 & 0.035 & $-21.747$ & 0.448 & 2.20 & 0.20 & 0.09 & spiral \\
A2589-00006 & 23:22:37.1 & +16:29:19 & 0.037 & $-21.706$ & 0.818 & 1.36 & 0.55 & 0.72 & elliptical \\
A2589-00007 & 23:23:59.1 & +16:48:40 & 0.043 & $-21.654$ & 0.863 & 0.39 & 0.03 & 0.91 & spiral (bulge-dominated) \\
A2589-00008 & 23:23:53.5 & +16:52:48 & 0.045 & $-21.653$ & 0.875 & 1.17 & 0.09 & 0.30 & disky elliptical \\
A2589-00010 & 23:23:56.8 & +16:44:60 & 0.038 & $-21.500$ & 0.837 & 1.01 & 0.07 & 0.68 & elliptical \\
A2589-00011 & 23:25:55.6 & +17:29:21 & 0.046 & $-21.466$ & 0.591 & 1.55 & 1.03 & 0.03 & spiral \\
A2589-00012 & 23:25:13.4 & +16:24:26 & 0.039 & $-21.427$ & 0.856 & 0.78 & 0.63 & 0.40 & disky elliptical \\
A2589-00013 & 23:24:05.8 & +16:47:31 & 0.042 & $-21.426$ & 0.879 & 0.09 & 0.06 & 0.73 & edge-on disk \\
A2589-00014 & 23:23:55.8 & +16:55:00 & 0.041 & $-21.390$ & 0.853 & 0.06 & 0.13 & 0.62 & disky elliptical \\
A2589-00015 & 23:23:02.1 & +17:32:20 & 0.038 & $-21.381$ & 0.775 & 1.24 & 0.93 & 0.69 & barred? spiral \\
A2589-00018 & 23:23:54.4 & +16:40:50 & 0.045 & $-21.321$ & 0.828 & 1.24 & 0.16 & 0.68 & elliptical \\
\enddata
\label{galinfo}
\end{deluxetable*}

\subsection{Morphological Classification}\label{morcla}

\begin{figure*}
\centering
\includegraphics[width=0.95\textwidth]{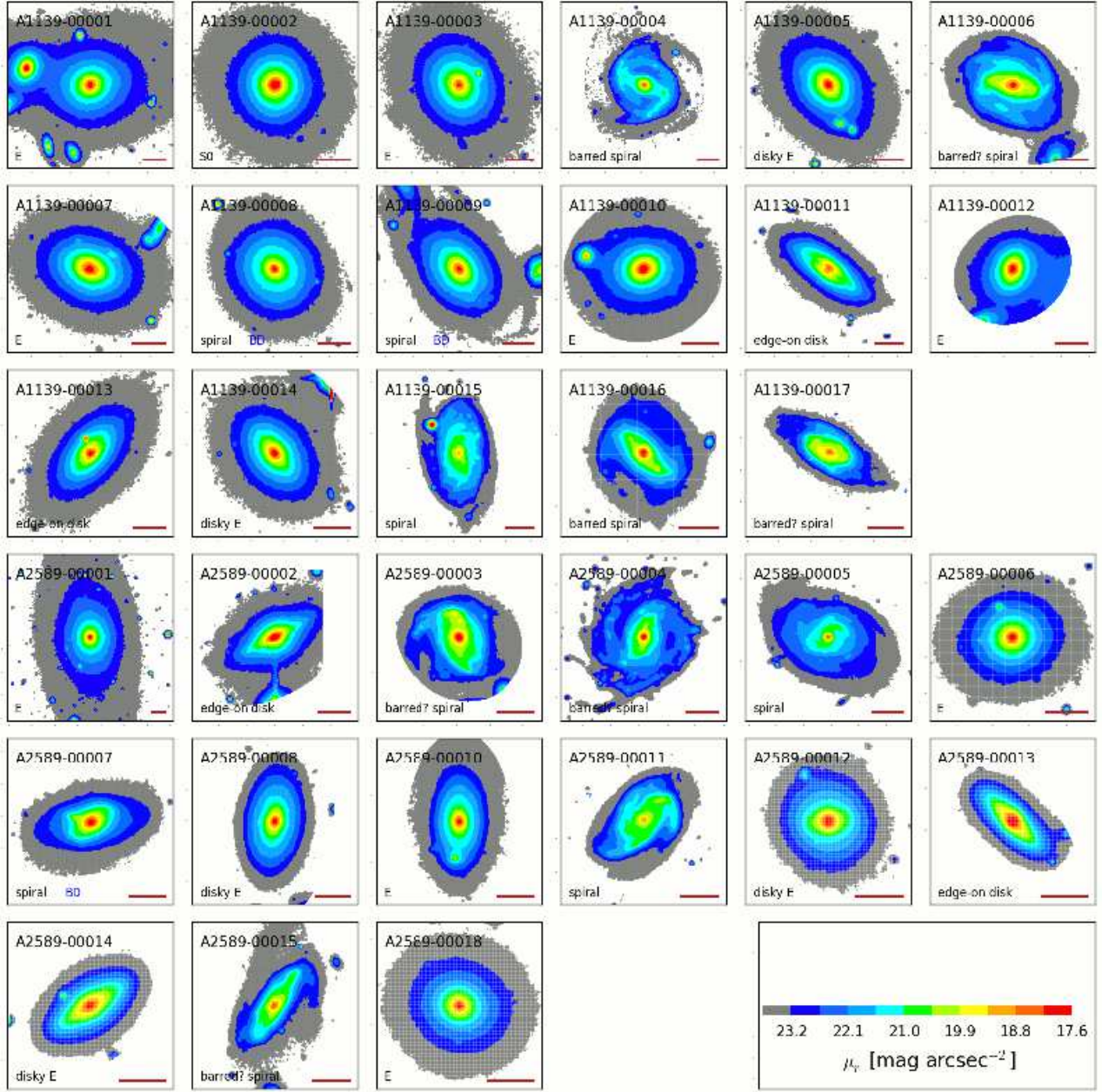}
\caption{The $r$-band surface brightness maps of the sample galaxies. Visually classified morphology is denoted at the lower left corner and the 10 arcsec ($\sim8$ kpc) scale bar is shown at the lower right corner in each panel. See the main text for the detailed description of each morphological type. \label{mumap}}
\end{figure*}

\begin{figure*}
\centering
\includegraphics[width=0.95\textwidth]{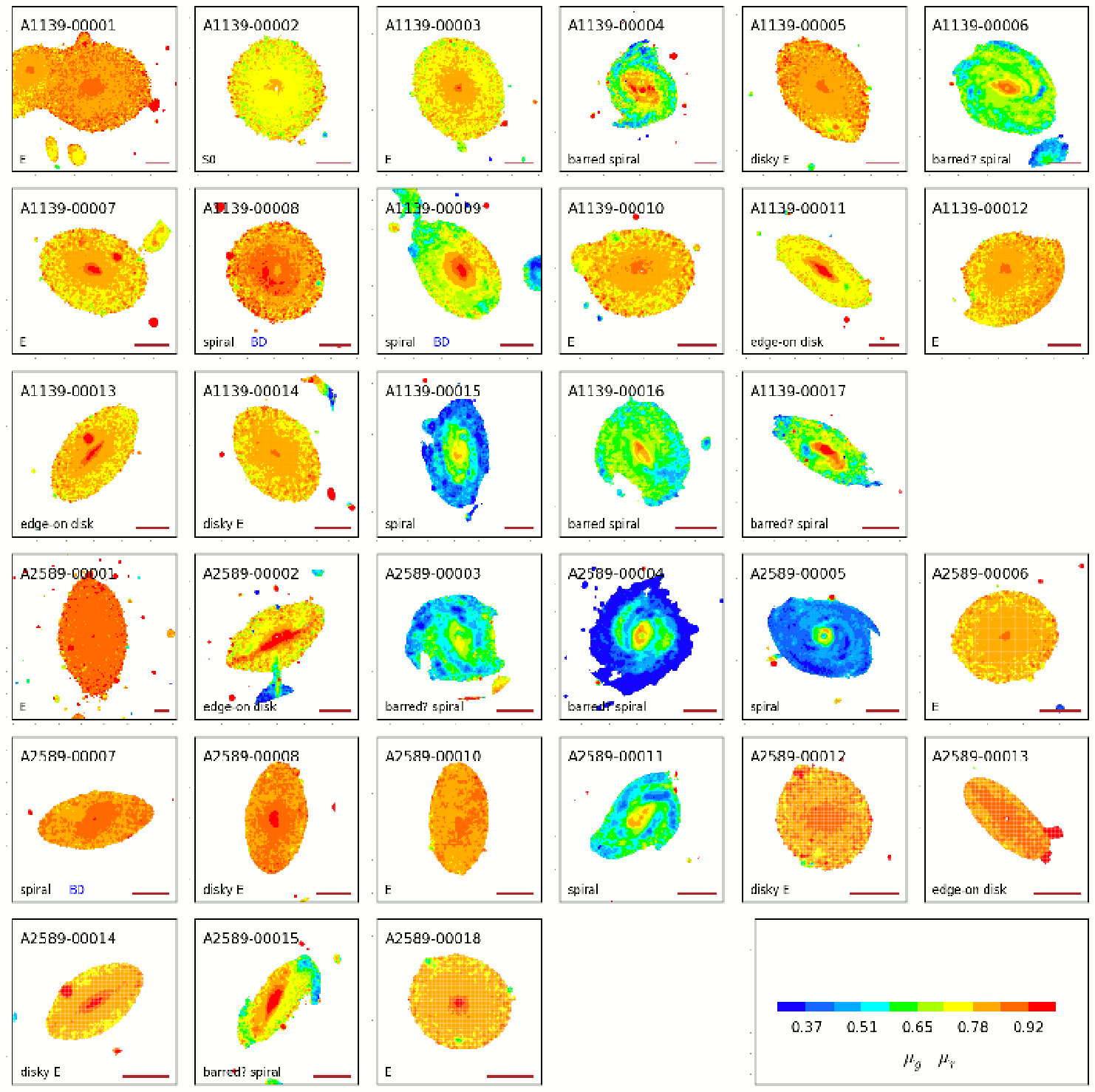}
\caption{The $g-r$ color maps of the sample galaxies. The pixels fainter than $\mu_r = 23.2$ {\umu} are not plotted. The visually classified morphology and the 10 arcsec scale bar are also denoted.\label{colmap}}
\end{figure*}

Figure~\ref{mumap} shows the $r$-band surface brightness ($\mu_r$) maps, which reveal the internal structures of our sample galaxies more clearly. Based on Figures~\ref{portrait} and \ref{mumap}, JHL, MP and HRL visually classified the morphological types of our sample galaxies into the following classes:
\begin{itemize}
 \item Typical elliptical (E): well-defined elliptical structures without any disturbed features,
 \item Disky elliptical (disky E): overall elliptical morphology with slightly disky internal structures (but not significantly different from typical elliptical galaxies),
 \item Lenticular (S0): hosting a faint disk component and not significantly inclined (face-on),
 \item Edge-on disk: largely inclined S0 or late-type (but mostly bulge-dominated in our sample),
 \item Barred spiral: spiral arms connected to a bar structure (a question mark is denoted when the bar is ambiguous), and
 \item Spiral: spiral arms without a bar structure.
\end{itemize}
Since the bulge fraction affects the overall properties of a late-type galaxy, a mark of `BD' is additionally noted for a bulge-dominated late-type galaxy.
The morphological classification is sometimes ambiguous: E versus disky E; disky E versus S0; and S0 versus bulge-dominated spiral. However, the small ambiguity between some types does not significantly matter in this paper, because this paper is mainly focused on the difference between broadly divided early- and late-type galaxies rather than between individual fine classes.
Note that the classification in this paper was conducted independently of that in the KYDISC catalog, and thus they may not necessarily coincide with each other. For example, A1139-00008 is listed as an S0 galaxy in the KYDISC catalog, but we classify it into a bulge-dominated spiral, because of very faint ambient features around it.
As a quantitative indicator of morphology, we also estimated the bulge-to-total ratio (B/T) of each galaxy, based on the decomposition into Sersic component (bulge) + exponential component (disk) using the GALFIT code \citep{pen02,pen10}.

Figure~\ref{colmap} presents the $g-r$ color maps of our sample galaxies. The internal distributions of stellar populations are revealed in these maps, which are particularly useful to visually inspect the interactions with neighbor galaxies (e.g., A1139-00009 and A1139-00017). Note that our morphological classification is not based on these color maps, but only based on the light distributions in the $r$ band.

\subsection{WISE Data}\label{wisedata}

Infrared color is a useful tool to diagnose star formation and galactic nuclei activities of galaxies \citep{ass10,jar11,jar17,ko16}.
In this paper, we use the Wide-field Infrared Survey Explorer \citep[WISE;][]{wri10} data. 
The WISE mapped the whole sky in four infrared bands: 3.4, 4.6, 12 and 22 $\mu$m (W1, W2, W3 and W4) with angular resolutions of 6.1$''$, 6.4$''$, 6.5$''$, and 12.0$''$, respectively. 

From the WISE website\footnote{http://wise2.ipac.caltech.edu/docs/release/allsky/}, we retrieved infrared magnitudes for our target galaxies. For most point-like sources in the WISE data, the recommended magnitude type is the \emph{profile-fit magnitude}, but we use the \emph{elliptical aperture magnitude} instead, because our targets are mostly extended sources even in the WISE images. The elliptical apertures in the WISE photometry are based on the cross-matching with  the Two Micron All Sky Survey Extended Source Catalog \citep[2MASS XSC;][]{skr06}.
However, the aperture magnitude is not available for one of our targets (A2589-00014), which means that this object looks like a point source in the infrared images although it is an extended source in higher resolution. Thus, we use the profile-fit magnitudes only for A2589-00014. 

We adopt the [4.6] $-$ [12] (W2 $-$ W3) color as a star formation indicator, because the [12] band flux is sensitive to star formation activity whereas the [4.6] band flux reflects age-independent stellar mass. On the other hand, the [3.4] $-$ [4.6] (W1 $-$ W2) color is known to be an indicator of active galactic nuclei (AGNs), owing to its sensitivity to hot dust \citep{jar17}. These color indices are used to describe the star formation properties of the sample galaxies.

\section{ANALYSIS}\label{anal}

\subsection{Standard Procedure for Masking Contaminants}\label{standard}

\begin{figure*}[!ht]
\centering
\includegraphics[width=0.95\textwidth]{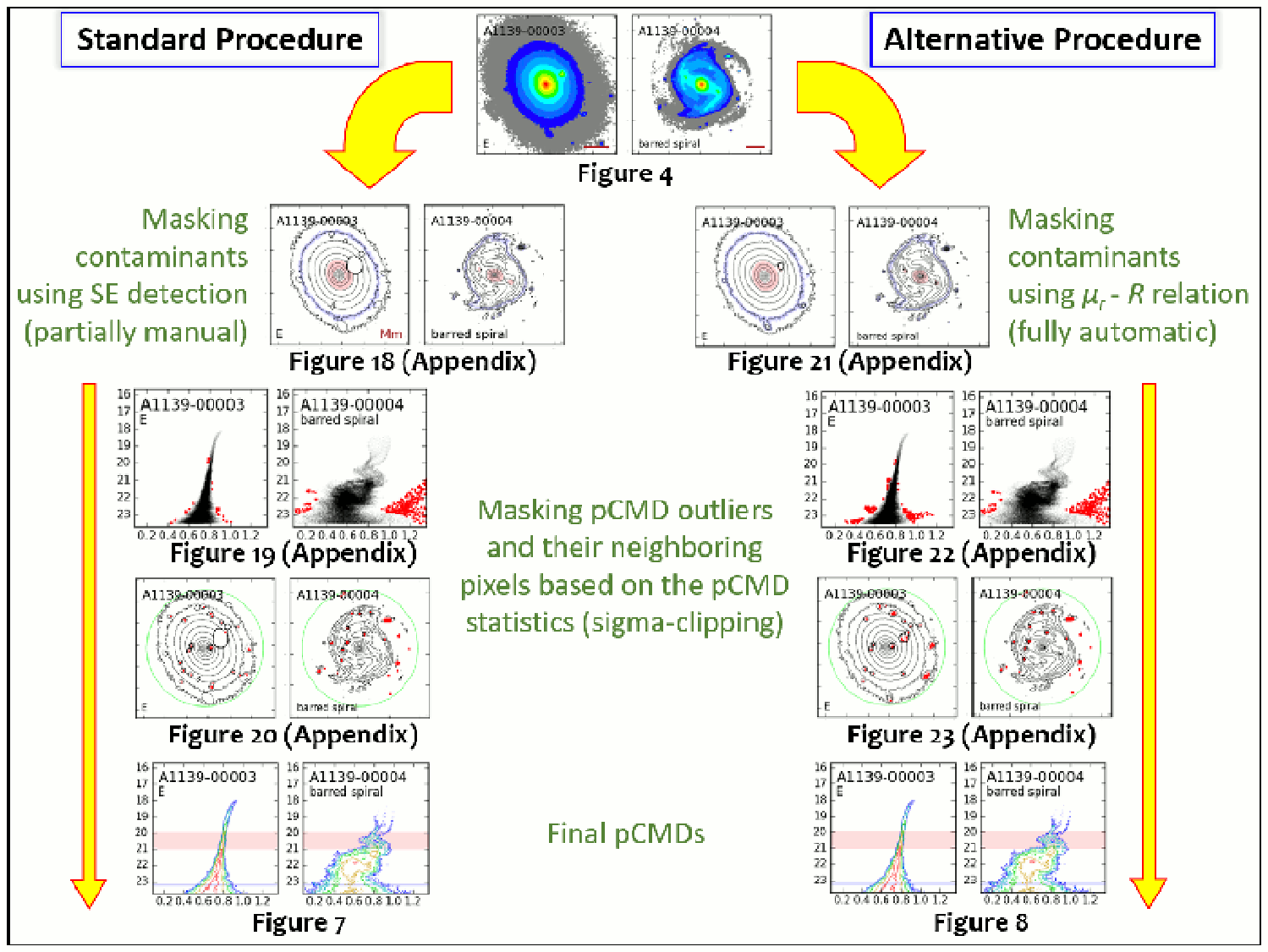}
\caption{Schematic summary of the standard and alternative procedures. The left-side sequence shows the standard procedure (described in Section~\ref{standard}), while the right-side sequence shows the alternative procedure (described in Section~\ref{altproc}). The full plots for the whole sample galaxies in each masking process are presented in Appendix~\ref{app1}.\label{procscheme}}
\end{figure*}

\begin{figure*}
\centering
\includegraphics[width=0.95\textwidth]{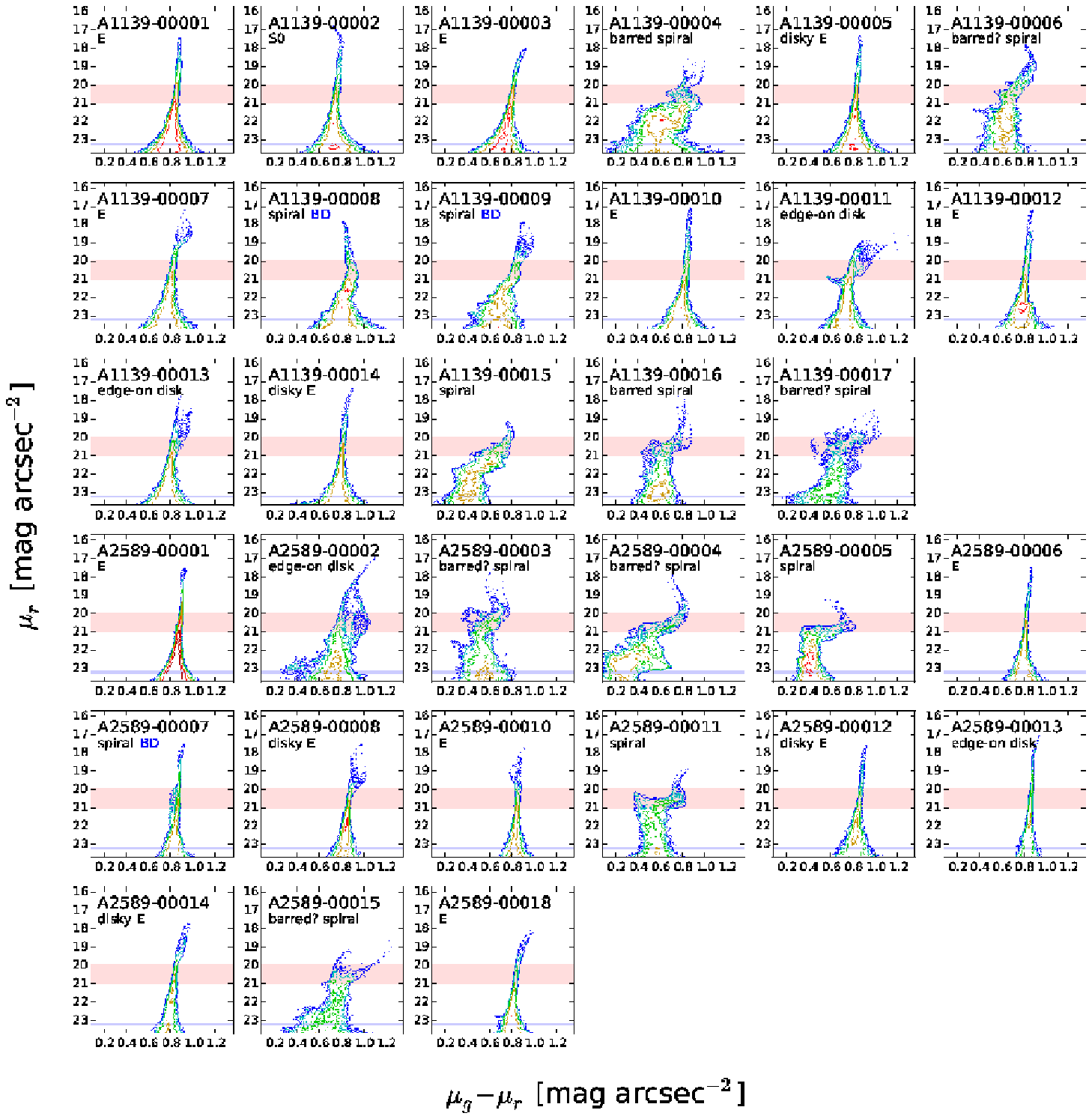}
\caption{The log-scale pixel number density contours of the final pCMDs after the standard masking procedure. The $20.0\le\mu_r\le21.0$ {\umu} range (faint red stripe) and the $\mu_r=23.2$ {\umu} limit (faint blue line) are denoted.\label{pcmdcon1}}
\end{figure*}

\begin{figure*}
\centering
\includegraphics[width=0.95\textwidth]{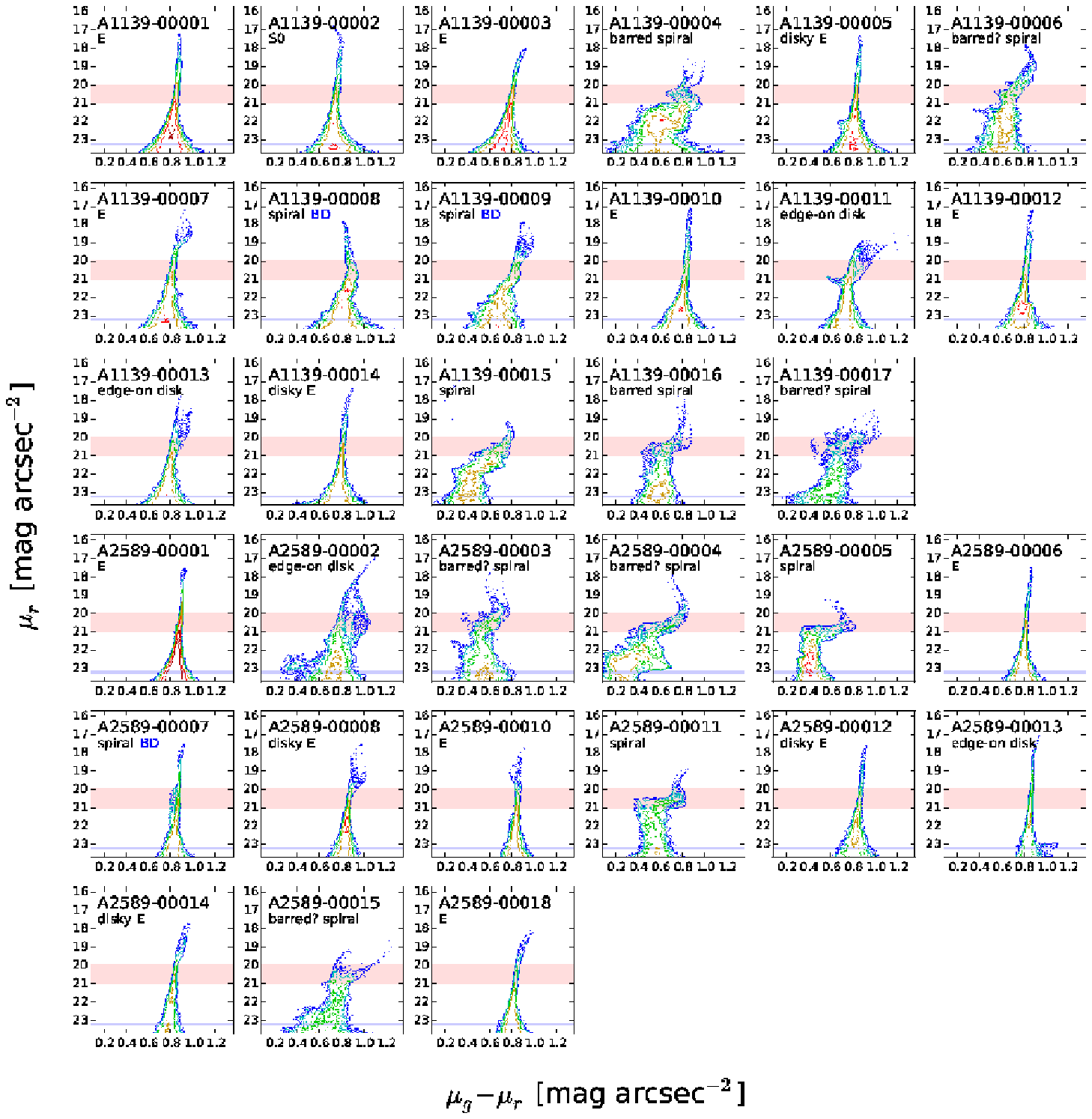}
\caption{The log-scale pixel number density contours of the final pCMDs after the alternative masking procedure. The $20.0\le\mu_r\le21.0$ {\umu} range (faint red stripe) and the $\mu_r=23.2$ {\umu} limit (faint blue line) are denoted. \label{pcmdcon2}}
\end{figure*}

\begin{figure*}
\centering
\includegraphics[width=0.95\textwidth]{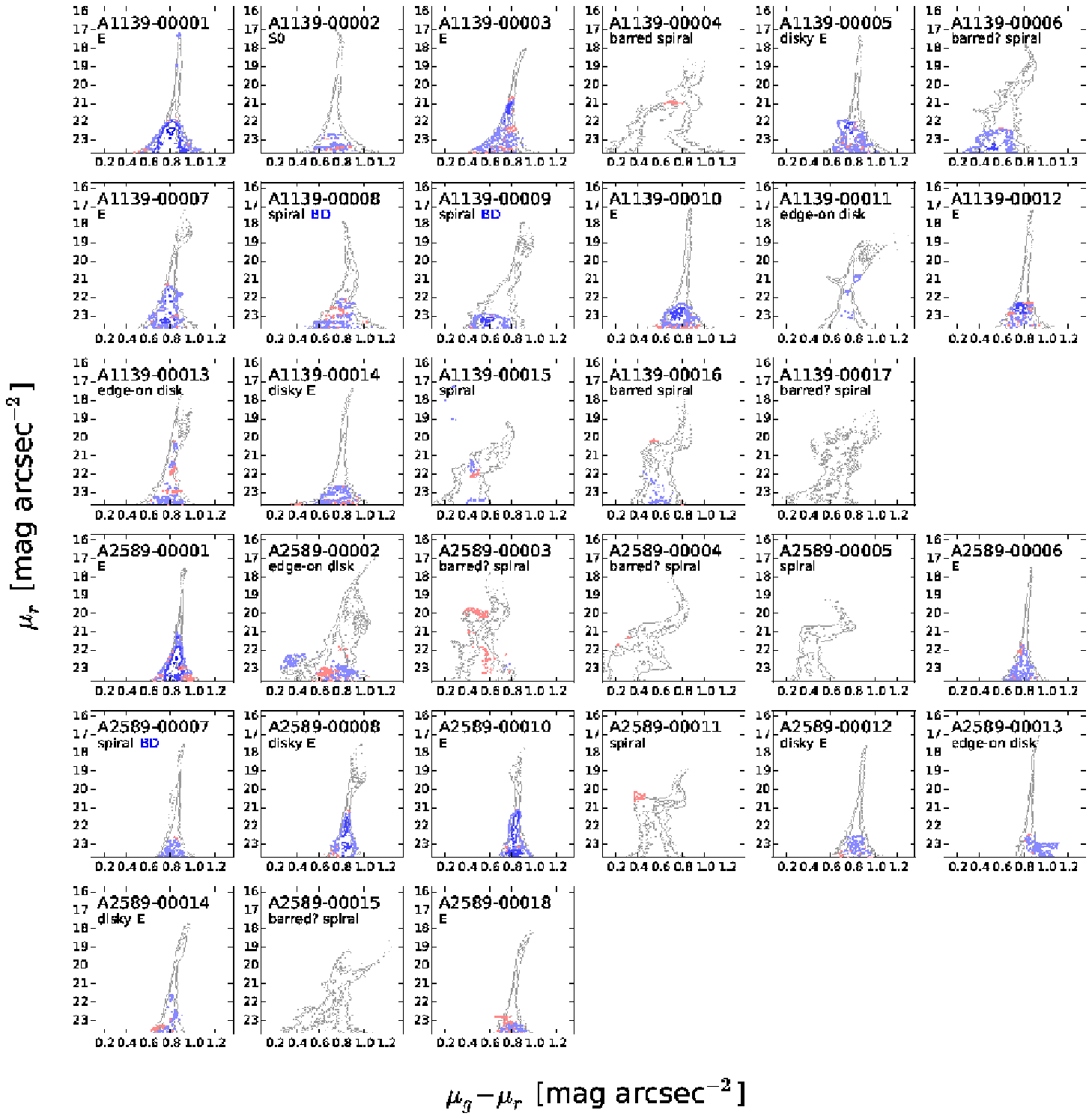}
\caption{The log-scale contours of the difference in pixel number density of pCMDs between the two masking procedures (Figure~\ref{pcmdcon1} subtracted by Figure~\ref{pcmdcon2}). The red contours indicate the domain at which the pCMDs after the standard masking procedure are denser (Figure~\ref{pcmdcon1} $>$ Figure~\ref{pcmdcon2}), whereas the blue contours show the opposite cases (Figure~\ref{pcmdcon1} $<$ Figure~\ref{pcmdcon2}).
The gray contours show the pCMDs after the standard masking procedure. \label{dpcmd}}
\end{figure*}

Before plotting final pCMDs, target images need to be appropriately processed, to minimize the influence of contaminants and to improve the signal-to-noise ratio of each pixel.
\citet{lee17} established a procedure for the pCMD analysis, which includes contaminants-masking and pixel-smoothing. The procedure is summarized as follows: 
\begin{enumerate}
 \item Trim a sufficient area around a target galaxy,
 \item Detect objects using the Source Extractor \citep[SE;][]{ber96} with sufficiently large background mesh sizes (to detect relatively large contaminants),
 \item Use the SE again with small background mesh sizes (to detect small and faint contaminants)
 \item Mask the pixels within the apertures of any detected non-target objects, and
 \item Smooth pixels by adopting a smoothing kernel with a seeing-sized ($0.8''$) aperture.
\end{enumerate}
In \citet{lee17}, this procedure worked well, because the target galaxies were BCGs with relatively simple shapes and little complex substructures. 

In this paper, on the other hand, the targets are galaxies in variety of morphology, from elliptical to spiral. Because elliptical galaxies (and most edge-on disk galaxies) do not show complex structures in their $r$-band images, there is no problem in applying the SE-detection-based masking to them. In the case of late-type galaxies, however, it does not work well to determine the masking apertures using the SE, because the SE hardly distinguish between contaminants (companions or foreground/background objects) and galaxy substructures (spiral arms, star-forming clumps, and bars). For a few late-type galaxies, the SE-detection-based masking removes more than $80\%$ of the entire image for a target galaxy, which is obviously over-masking.

Thus, we apply the SE-detection-based masking adaptively according to the target morphology and the masking performance. That is, the full procedure is applied if the SE-detection-based masking does not cause over-masking (e.g., A1139-00003). On the other hand, if the SE-detection-based masking (step 2 and/or step 3 in the procedure) appears to over-mask the target galaxy, the masking step is omitted (e.g., A1139-00004).
During this adaptation, we intended to minimize the omission of masking. That is, the masking step is omitted only when it obviously over-masks the target galaxy, whereas the masking is conducted when it is unclear whether the target is over-masked or not. At least among our sample galaxies, there was rarely ambiguous situation and thus we decided the final masking sets relatively at ease. 

However, the adaptive masking has a fundamental weakness: late-type galaxies with complicated structures tend to be masked too little, compared to early-type galaxies with simple structures. Because late-type galaxies are expected to have remaining contaminants that were not sufficiently masked in the SE-detection-based masking process, a supplementary masking process is necessary.
For this, we additionally masked the pixels that correspond to the pCMD outliers. The processes are listed as follows:
\begin{enumerate}
 \item Estimate the 5, 50, and 95 percentiles in the pixel $g-r$ color distribution at given $\mu_r$, using the pCMD after the SE-detection-based masking,
 \item Define $\sigma_1\equiv (g-r)_{50\%} - (g-r)_{5\%}$ and $\sigma_2\equiv (g-r)_{95\%} - (g-r)_{50\%}$ as a function of $\mu_r$,
 \item Define pCMD outliers as the pixels with $g-r<(g-r)_{50\%} - 2 \sigma_1$ or $g-r>(g-r)_{50\%} + 2 \sigma_2$, at given $\mu_r$, and
 \item Mask the pCMD outliers and their neighboring pixels within $0.8''$ (seeing size).
\end{enumerate}

Figure~\ref{procscheme} schematically summarizes the entire masking procedures to yield the final pCMDs (the standard procedure at the left side).
The plots at the second row of the left side in Figure~\ref{procscheme} show the surface brightness contour maps of the sample galaxies after the adaptive (SE-detection-based) masking.
The plots at the third row of the left side show the pCMDs after the SE-detection-based masking and the smoothing with the $0.8''$-aperture spline kernel \citep{lee17}, with the pCMD outliers marked (see Appendix~\ref{app1} for the full plots of the whole sample). The pCMD outliers of late-type galaxies mostly outnumber those of early-type galaxies, which means that many remaining contaminants in late-type galaxies are additionally masked in this process.

Note that the pCMD outlier masking is not perfect either. There may be still remaining contaminants or to the contrary there may be some over-masking particularly for fine substructures in the target galaxies. For example, the BCG of A2589 (A2589-00001) is thought to have vestiges of infalling low-mass star-forming satellites, which was revealed from the analysis of its pCMD outliers in \citet{lee17}. Such fine features are mostly washed out in the pCMD outlier masking process.
Thus, this process can not be used if we want to inspect the fine substructures in target galaxies. However, it does not significantly affect the main features of the pCMD backbones, because the number of the outliers are typically much smaller than that of the pixels in main features.
The plots at the fourth row in Figure~\ref{procscheme} present the $\mu_r$ contour maps after masking pCMD outliers and their neighboring pixels within $0.8''$. After these two-step masking processes, the final pCMDs are yielded as shown in Figure~\ref{pcmdcon1}.

\subsection{Alternative Procedure}\label{altproc}

\begin{deluxetable*}{ll}
\tabletypesize{\footnotesize}
\tablenum{2} \tablecolumns{2} \tablecaption{Parameters Describing pCMD Features} \tablewidth{0pt}
\tablehead{ pCMD Parameter & Description}
\startdata
$\mu_r$(tip) & Minimum (brightest) $r$-band surface brightness among all pixels \\
min$(g-r)_{\mu_r\le23.2}$ & Minimum (bluest) value of mean $g-r$ color at $\mu_r\le23.2$ {\umu} \\
max$(g-r)_{\mu_r\le23.2}$ & Maximum (reddest) value of mean $g-r$ color at $\mu_r\le23.2$ {\umu} \\
min$(\sigma(g-r))_{\mu_r\le21.2}$ & Minimum (tightest) value of $g-r$ color dispersion at $\mu_r\le21.2$ {\umu} \\
max$(\sigma(g-r))_{\mu_r\le21.2}$ &  Maximum (most dispersed) value of $g-r$ color dispersion at $\mu_r\le21.2$ {\umu} \\
max$(\sigma(g-r))_{20.0\le\mu_r\le21.0}$ & Maximum value of $g-r$ color dispersion at $20.0\le\mu_r\le21.0$ {\umu} \\
max$(\sigma(g-r))_{18.5\le\mu_r\le19.5}$ & Maximum value of $g-r$ color dispersion at $18.5\le\mu_r\le19.5$ {\umu} \\
\enddata
\tablecomments{The `mean $g-r$ color' and `$g-r$ color dispersion' indicate the quantities estimated using the pixels with fixed $\mu_r$. That is, min$(g-r)_{\mu_r\le23.2}$ is the bluest value among the `mean $g-r$ colors as a function of $\mu_r$', not the $g-r$ color of the bluest pixel among the whole pixels. In the text, `the minimum (maximum) $g-r$ color' indicates `the minimum (maximum) value of mean $g-r$ color as a function of $\mu_r$'.}
\label{params}
\end{deluxetable*}

In the standard procedure, the SE-detection-based masking process for contaminants has two weaknesses. One is that late-type galaxies can not be sufficiently masked using the SE-detection-based method, which is supplemented to some extent using the pCMD outlier masking as described in Section~\ref{standard}.
However, another weakness still remains: the `adaptive masking'  can cause a bias when we compare the pCMD properties between different morphological types.
As already mentioned, late-type galaxies tend to be less suitable for the SE-detection-based masking than early-type galaxies. Among the 13 spiral galaxies, the SE-detection-based masking is completely omitted for four spirals (A1139-00004, A2589-00003, A2589-00004, and A2589-00011), while the full masking processes are applied to only two spirals (A1139-00008 and A2589-00007). On the other hand, the full masking processes were applied to every elliptical galaxy. That is, the masking processes for early- and late-type galaxies are not identical in the standard procedure, and thus the comparison of the final pCMDs between galaxies with different morphological types may not be fair.

To address this issue, we try an alternative masking procedure that does not depend on subjective `adaptation' according to galaxy morphology. This procedure is based on the idea that the light from contaminants tends to be brighter than the brightness expected at the distance from the target galaxy center. The detailed procedure is as follows:
\begin{enumerate}
 \item Trim a sufficient area around a target galaxy,
 \item Trim the image again with a radius of $f_c\times R80_{(22.2-23.2)}$, where $f_c$ is a fixed value and $R80_{(22.2-23.2)}$ is the 80 percentile in the distribution of distance to center, among pixels with $22.2\le\mu_r\le23.2$ {\umu}; we empirically choose $f_c=1.5$ (see Appendix~\ref{app1} for the trimming area for each sample galaxy),
 \item Estimate the 10 and 50 percentiles in the distribution of distance to center, among pixels with given $\mu_r$  ($R10_{\mu_r}$ and $R50_{\mu_r}$, respectively),
 \item Mask pixels with $\mu_r$ and radial distance ($R$) satisfying $R - R10_{\mu_r}> f_R\times (R50_{\mu_r} - R10_{\mu_r})$, where we empirically choose $f_R=5$, and
 \item Conduct the additional pCMD outlier masking as described in Section~\ref{standard}.
\end{enumerate}
These processes depend on a few control parameters empirically determined, which is a weakness. Nevertheless, this method has a considerable merit: it can be consistently applied to target galaxies regardless of their morphological types.

The plots at the second row of the right side in Figure~\ref{procscheme} show the surface brightness contour maps after this $\mu_r$-radius-relation-based ($R(\mu_r)$-based) masking. The overall area masked by the $R(\mu_r)$-based method tends to be smaller (under-masking) than that by the SE-detection-based method.
Thus, the role of the pCMD outlier masking is more important in the alternative procedure, because it supplements the weak performance of the $R(\mu_r)$-based masking.
In this procedure, on the other hand, the number of pCMD outliers is averagely not so different between early- and late-type galaxies, unlike that in the standard procedure. This is because the alternative masking procedure is free from the morphology bias caused by subjective `adaptation'. 

The final pCMDs after the alternative masking procedure are presented as pixel number density contours in Figure~\ref{pcmdcon2}. The overall features of the pCMDs in Figure~\ref{pcmdcon2} are not significantly different from those after the standard masking procedure in Figure~\ref{pcmdcon1}, but some small details appear to be in disagreement. Such differences are revealed in Figure~\ref{dpcmd}, which shows the pixel number density contours for the pCMD residuals (Figure~\ref{pcmdcon1} subtracted by Figure~\ref{pcmdcon2}). In most cases, the pixels in the pCMDs after the alternative procedure outnumber those after the standard procedure (blue contours). This means that the alternative procedure is less efficient in masking contaminants than the standard procedure at least for early-type galaxies.
On the other hand, the opposite case is much rarer (red contours) even for late-type galaxies.
It is hard to increase the masking efficiency of the alternative procedure by adjusting $f_R$, because a too small $f_R$ value frequently results in unreasonable masking.

In summary, the alternative procedure is not sufficiently good in the aspect of masking efficiency, but the results from the alternative procedure are necessary when technically fair comparison without any morphology bias is required.
In Section~\ref{result}, we mainly analyze the results from the standard procedure, to take an advantage of its masking efficiency (mainly for early-type galaxies). However, the results from the alternative procedure are also used to check the effect of morphology bias, which are fully presented in Appendix~\ref{app2}.

\section{RESULTS}\label{result}

\begin{figure*}
\centering
\includegraphics[width=0.95\textwidth]{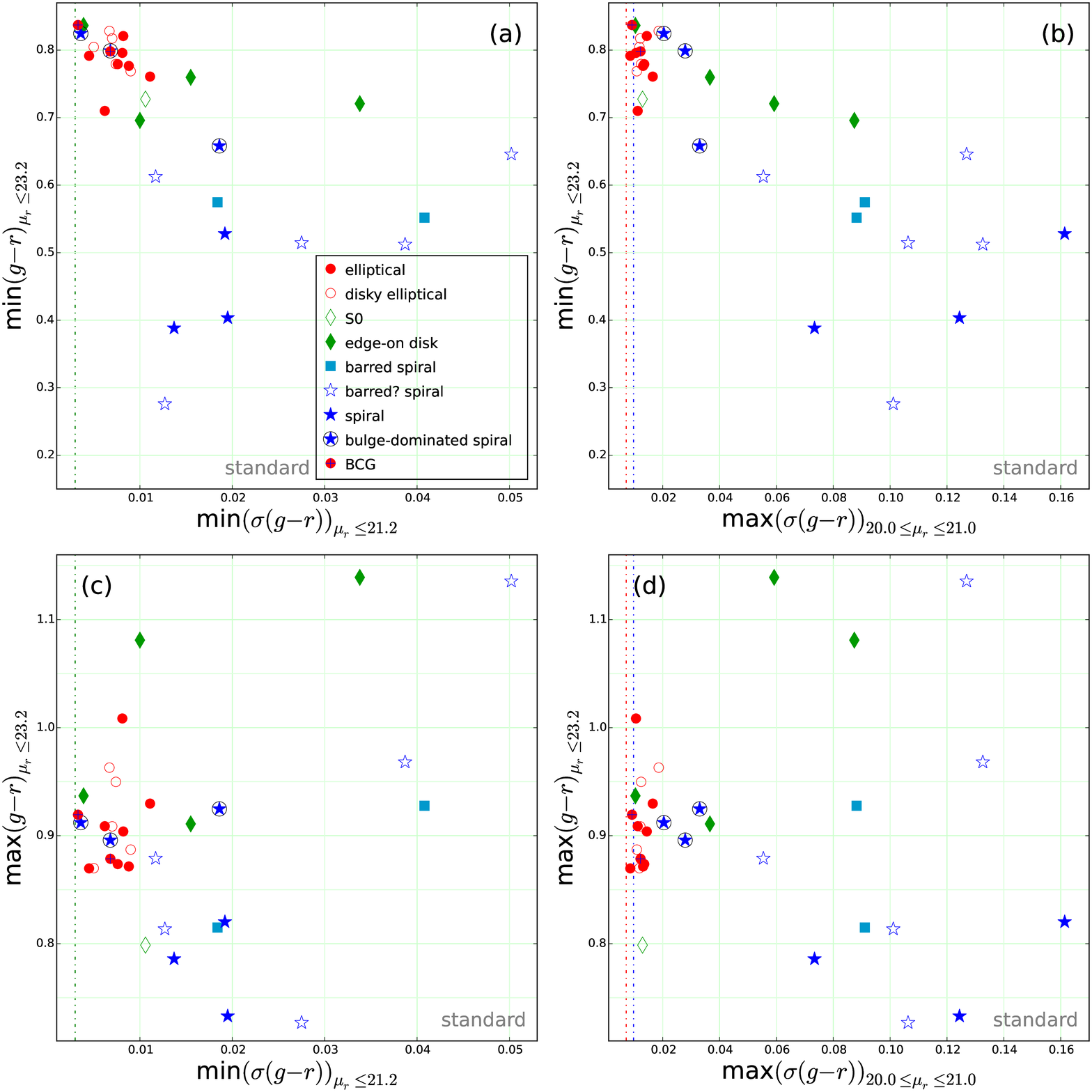}
\caption{Correlations between pCMD parameters, based on the pCMDs after the standard procedure:
(a) the minimum $g-r$ color at $\mu_r\le23.2$ {\umu} versus the minimum $g-r$ color dispersion at $\mu_r\le21.2$ {\umu},
(b) the minimum $g-r$ color at $\mu_r\le23.2$ {\umu} versus the maximum $g-r$ color dispersion at $20.0\le\mu_r\le21.0$ {\umu},
(c) the maximum $g-r$ color at $\mu_r\le23.2$ {\umu} versus the minimum $g-r$ color dispersion at $\mu_r\le21.2$ {\umu}, and
(d) the maximum $g-r$ color at $\mu_r\le23.2$ {\umu} versus the maximum $g-r$ color dispersion at $20.0\le\mu_r\le21.0$ {\umu}. The dot-dashed lines represent the lower limits of color dispersion, above which color dispersion is dominated by intrinsic color scatter rather than photometric uncertainty, at 21.0 {\umu} for A1139 (blue; in panels (b) and (d)), at 21.0 {\umu} for A2589 (red; in panels (b) and (d)), and at 19.2 {\umu} for both clusters (green; in panels (a) and (c)).\label{coldev1}}
\end{figure*}

\begin{figure*}[!ht]
\centering
\includegraphics[width=0.95\textwidth]{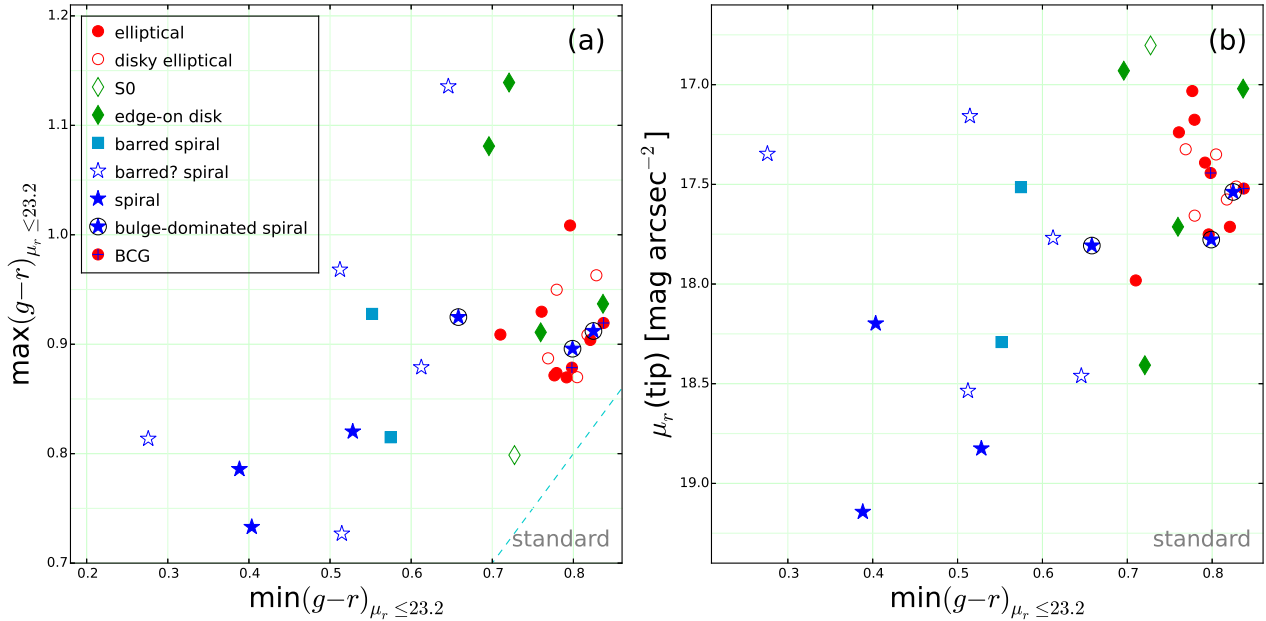}
\caption{(a) The maximum versus minimum values of mean $g-r$ color at $\mu_r\le23.2$ {\umu}, and (b) the brightest $\mu_r$ versus the minimum $g-r$ color at $\mu_r\le23.2$ {\umu}, based on the pCMDs after the standard masking procedure. Various symbols indicate the morphological types, and the dashed line (cyan) shows the one-to-one relation. \label{stat1}}
\end{figure*}

\begin{figure*}
\centering
\includegraphics[width=0.95\textwidth]{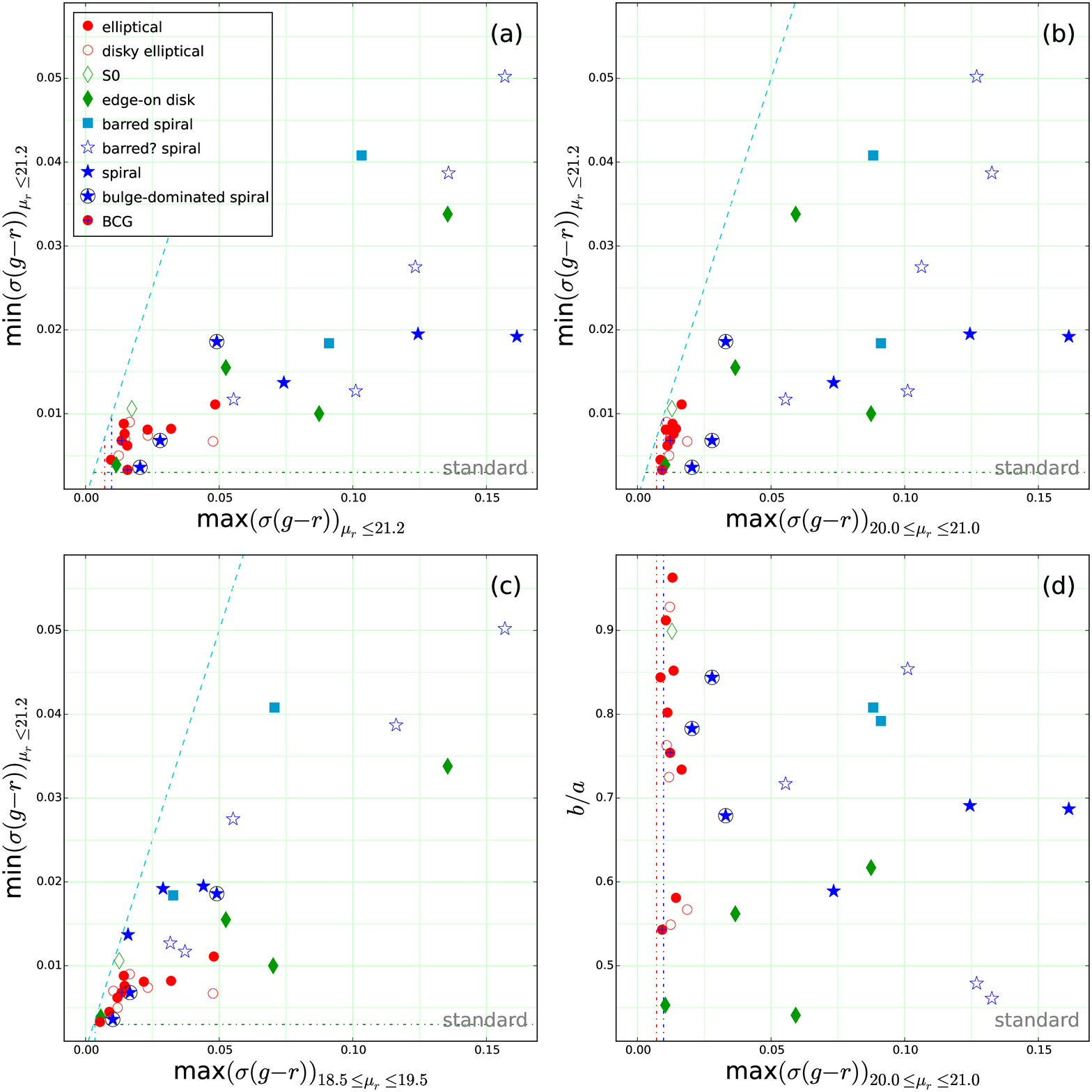}
\caption{The minimum $g-r$ color dispersion at $\mu_r\le21.2$ {\umu} (a) versus the maximum $g-r$ color dispersion at $\mu_r\le21.2$ {\umu}, (b) versus the maximum $g-r$ color dispersion at $20.0\le\mu_r\le21.0$ {\umu}, and (c) versus the maximum $g-r$ color dispersion at $18.5\le\mu_r\le19.5$ {\umu}. (d) Axis ratio versus the maximum $g-r$ color dispersion at $20.0\le\mu_r\le21.0$ {\umu}. The dashed lines (cyan) show the one-to-one relations. The blue, red and green dot-dashed lines are the reliability limits as described in Figure~\ref{coldev1}, and the cyan vertical dot-dashed line in Panel (c) is the lower limit at 19.5 {\umu} for both clusters. All parameters are based on the pCMDs after the standard masking procedure. \label{class1}}
\end{figure*}

\begin{figure*}[!ht]
\centering
\includegraphics[width=0.95\textwidth]{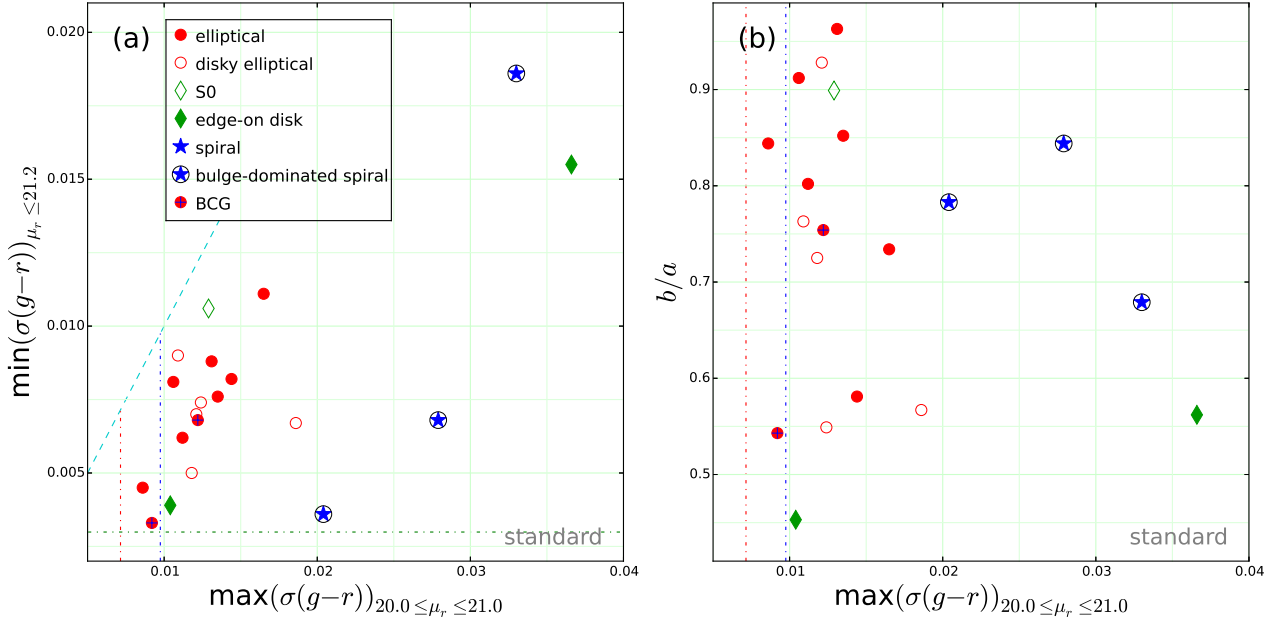}
\caption{Zoom-in plots of (a) the minimum $g-r$ color dispersion at $\mu_r\le21.2$ {\umu} versus the maximum $g-r$ color dispersion at $20.0\le\mu_r\le21.0$ {\umu} (Figure~\ref{class1}(b)), and (b) axis ratio versus the maximum $g-r$ color dispersion at $20.0\le\mu_r\le21.0$ {\umu} (Figure~\ref{class1}(d)). Both of the plots are based on the pCMDs after the standard masking procedure.\label{classin}}
\end{figure*}

\begin{figure*}[!t]
\centering
\includegraphics[width=0.95\textwidth]{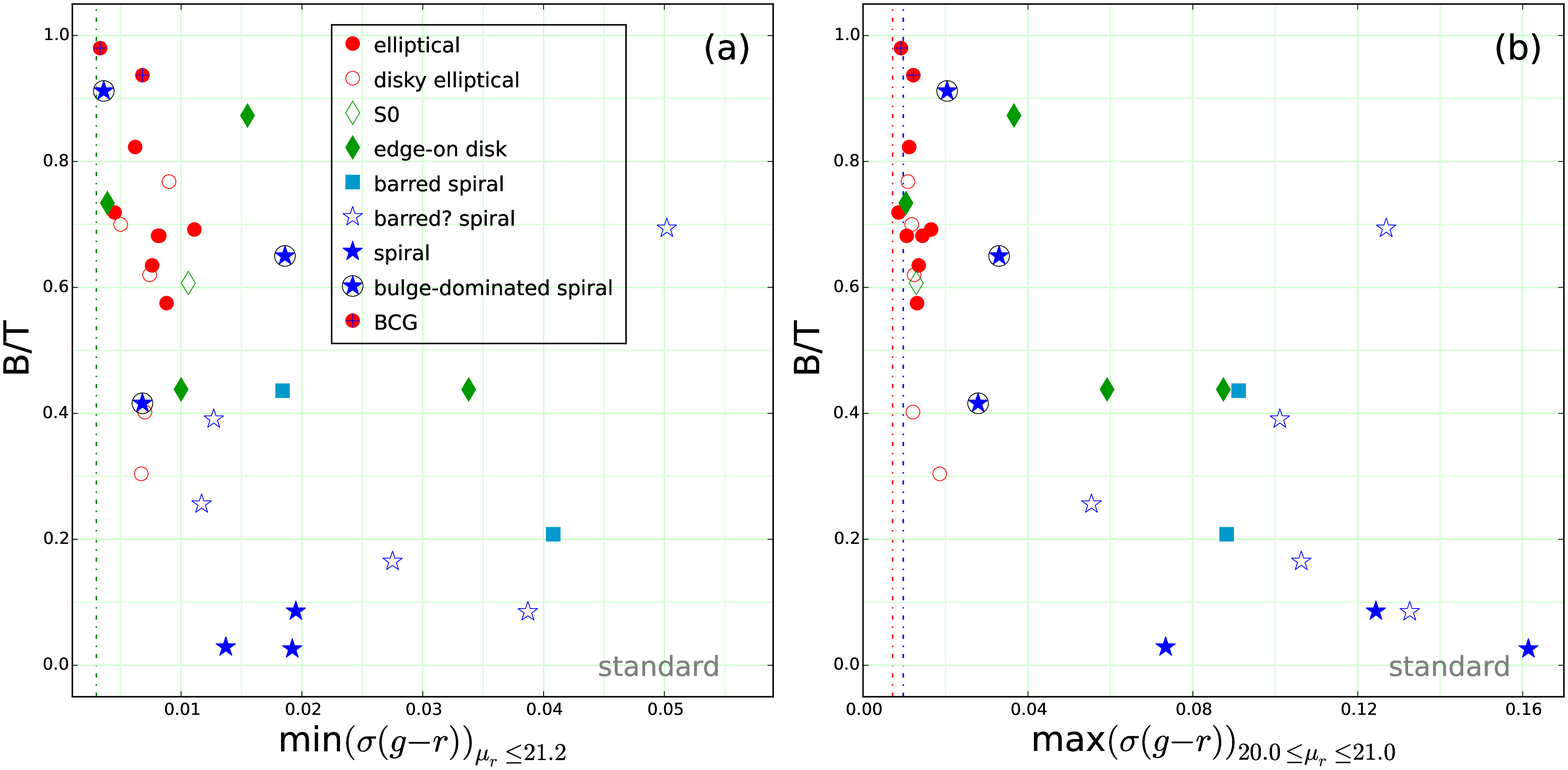}
\caption{ Comparison with bulge-to-total ratio (B/T) of (a) minimum $g-r$ color dispersion at $\mu_r\le21.2$ {\umu} and (b) maximum $g-r$ color dispersion at $20.0\le\mu_r\le21.0$ {\umu}. \label{btr}}
\end{figure*}

\subsection{Overall Description of the pCMDs}

In the final pCMDs (Figures~\ref{pcmdcon1} and \ref{pcmdcon2}), early-type galaxies show simple pCMD features, whereas late-type galaxies show significantly curved pCMDs. As introduced in Section~\ref{intro}, those features are called prime sequences and inverse-L features in \citet{lan07}, respectively. However, such a dichotomic division is not always obvious. Some early-type galaxies have considerably disturbed features at their bright parts (A1139-00007, A2589-00008 and A2589-00014), and besides the pCMDs of some bulge-dominated late-type galaxies or edge-on disk galaxies are similar to those of early-type galaxies (A2589-00007 and A2589-00013).
That is, galaxies even in similar morphological types have variety in their pCMD features, and such variety is larger in late-type galaxies.

In the case of early-type galaxies, the unusual features (deviated from the prime sequences) seem to be closely related to tidal interactions with nearby neighbors or companions. The prime sequences of the elliptical galaxies show curvatures different from one another, which implies that their formation histories are not entirely homologous as discussed in \citet{lee17}. Different internal dust extinction may be responsible for such variety, too \citep{lee11,lee12}. The only face-on S0 galaxy in our sample (A1139-00002) shows a blue pCMD-tip feature, which indicates that this galaxy may have experienced central star formation activity very recently.

The (barred and unbarred) spiral galaxies show complex and diverse features in their pCMDs. The pCMD structures of a spiral galaxy is roughly described to consist of a (red and bright) bulge part and a (blue and faint) disk part, which form an overall inverse-L feature. However, whereas this division is quite clear in some spiral galaxies (A1139-00016, A2589-0004 and A2589-00011), some other spiral galaxies show more complicated and amorphous features (A1139-00017 and A2589-00015). The bulge-dominated spiral galaxies (A1139-00008, A1139-00009 and A2589-00007) show relatively simple pCMD features compared to the other spiral galaxies, but more distorted than the pCMDs of typical elliptical galaxies.

Finally, some edge-on disk galaxies (A1139-00013 and A2589-00013) have pCMDs similar to those of elliptical galaxies, but some other edge-on disk galaxies do not (A1139-00011 and A2589-00002). The former edge-on disk galaxies have quite elliptical appearances in their $r$-band images (Figure~\ref{portrait}), but they were classified into edge-on disks because an edge-on dust lane is found (A1139-0013) or the morphology is too elongated to be an elliptical galaxy (A2589-00013). The latter edge-on disks are clearly different from elliptical galaxies in their appearances.

\subsection{Quantitative Comparison}

One of the main purposes of this paper is to establish a quantitative method using pCMDs to distinguish between galaxies with different morphological types.
For quantitative comparison of the pCMD features, we devised several parameters describing the features of pCMDs as listed in Table~\ref{params}.

\subsubsection{Reliability Limits}

Following \citet{lee17}, we limit our analysis to $\mu_r\le23.2$ {\umu} for the use of pixel color, at which the photometric uncertainty of a single pixel is smaller than 0.03 {\umu}.
On the other hand, the limit for the use of pixel color dispersion\footnote{In this paper, the word `color dispersion' indicates the standard deviation of pixel colors at given $\mu_r$. This quantity was called `color deviation' in \citet{lee17}, but we replace it by `color dispersion' for better understanding.} ($\sigma(g-r)$) is $\mu_r\le21.2$ {\umu}. At $\mu_r>21.2$ {\umu}, color dispersion tends to be significantly affected by the photometric uncertainty of individual pixels and thus cannot represent the intrinsic scatter of pixel color \citep[see Figure~11 of][]{lee17}.

When we compare color dispersion values, the consideration of photometric uncertainty is important even at $\mu_r\le21.2$ {\umu}, depending on situation. \citet{lee17} empirically showed that the lower limit of reliable color dispersion at given $\mu_r$ is 1.3 $\times$ photometric uncertainty of a single pixel in A1139 and A2589. 
For example, if a max$(\sigma(g-r))_{20.0\le\mu_r\le21.0}$ value is close to the value of 1.3 $\times$ typical photometric uncertainty at $\mu_r = 21.0$ {\umu}, then the measured max$(\sigma(g-r))_{20.0\le\mu_r\le21.0}$ value indicates the upper limit rather than the intrinsic color dispersion. In other words, we can regard the color dispersion to be intrinsic only when the value is sufficiently larger than 1.3 $\times$ photometric uncertainty.

This should be also considered when we compare min$(\sigma(g-r))_{\mu_r\le21.2}$. Although the minimum color dispersion usually corresponds to the color dispersion at the brightest $\mu_r$ ($\mu_r$(tip)) and the photometric uncertainty of the brightest pixel is typically tiny, the reliability limit needs to be counted if the measured color dispersion is also very small. Because the faintest surface brightness among the brightest pixels of our sample galaxies (i.e., maximum $\mu_r$(tip)) is about 19.2 {\umu}, we can safely regard the min$(\sigma(g-r))_{\mu_r\le21.2}$ value to be intrinsic if it is larger than 1.3 $\times$ photometric uncertainty at $\mu_r=19.2$ {\umu}.
In the subsequent investigation, all interpretations of the figures are based on these considerations.

\subsubsection{Galaxy Morphology and pCMD Parameters}\label{qcomp}

From now on, we investigate the relationship between the pCMD parameters and galaxy morphology, and its physical implication.
Figure~\ref{coldev1} presents the basic correlations between the pCMD parameters, with the morphological types denoted. The trends between the parameters in the whole sample are mostly not obvious, but Figure~\ref{coldev1}(b) shows a notable anti-correlation between min$(g-r)_{\mu_r\le23.2}$ and max$(\sigma(g-r))_{20.0\le\mu_r\le21.0}$. This anti-correlation appears to be closely related with galaxy morphology, in the context that elliptical galaxies tend to have larger min$(g-r)_{\mu_r\le23.2}$ and smaller max$(\sigma(g-r))_{20.0\le\mu_r\le21.0}$ than the galaxies with the other types.
This is a quantitative expression of the fact that the elliptical galaxies hardly have blue pixels and their pCMDs tend to be tightly bound.

A similar plot was shown in Figure 13 of \citet{lan07}, in which the mean pixel color and the mean pixel color dispersion were compared. Although \citet{lan07} found a correlation between those mean values, its scatter is larger than that in our Figure~\ref{coldev1}(b). This indicates that the minimum pixel color and the maximum pixel color dispersion are better indicators of galaxy morphology than the mean values. On the other hand, the minimum pixel color dispersion (Figure~\ref{coldev1}(a)) or the maximum pixel color (Figure~\ref{coldev1}(c)-(d)) seems to be less efficient even than the mean values. The reason is not difficult to figure out: even late-type galaxies may have small min$(\sigma(g-r))$ values owing to their bulges, and large max$(g-r)$ values due to significant dust extinction in their disks as well as bulges. 

In Figure~\ref{stat1}, we examine how well galaxy morphology is distinguished by pCMD colors and $\mu_r$(tip).
In Figure~\ref{stat1}(a), the elliptical galaxies are distributed in a small domain, whereas the spiral galaxies show a much wider distribution. Most spiral galaxies tend to have pixel colors bluer than those of early-type galaxies, but the max$(g-r)_{\mu_r\le23.2}$ values of some late-type galaxies are similar to those of early-type galaxies (A1139-00017 and A1139-00006) or even larger (A2589-00015).
On the other hand, min$(g-r)_{\mu_r\le23.2}$ seems to divide elliptical galaxies from spiral galaxies better than max$(g-r)_{\mu_r\le23.2}$ does, except for bulge-dominated spiral galaxies (A1139-00008 and A2589-00007) and edge-on disk galaxies. These are because max$(g-r)_{\mu_r\le23.2}$ represents the color of a late-type galaxy's bulge which is known to be similar to an elliptical galaxy in their properties \citep{dre87,fis08}\footnote{But, see also \citet{gad09}.}, whereas min$(g-r)_{\mu_r\le23.2}$ is strongly affected by their disk parts.
In Figure~\ref{stat1}(b), the $\mu_r$(tip) is brighter than 18.0 {\umu} for all elliptical galaxies in our sample. On the other hand, many spiral galaxies have fainter $\mu_r$(tip), but some spiral galaxies have $\mu_r$(tip) as bright as those of elliptical galaxies (A2589-00003 and A2589-00004). These results show that the combination of pCMD colors and $\mu_r$(tip) moderately divides early- and late-type galaxies, but not perfectly.

\begin{deluxetable*}{lr@{.}lr@{.}lr@{.}lr@{.}lr@{.}lr@{.}lr@{.}lr@{.}l}
\tablenum{3} \tablecolumns{17} \tablecaption{Correlations with the WISE [4.6] $-$ [12] Color} \tablewidth{0pt}
\tablehead{ pCMD Parameters & \multicolumn{4}{c}{All} & \multicolumn{4}{c}{E} & \multicolumn{4}{c}{E + disky E} & \multicolumn{4}{c}{Spirals} }
\startdata
min($g-r$)$_{\mu_r\le23.2}$ &  $-0$ & 87 & (0 & 000) & 0 & 03 & (0 & 935) & 0 & 19 & (0 & 506) & $-0$ & 62 & (0 & 052) \\
max($g-r$)$_{\mu_r\le23.2}$ &  $-0$ & 25 & (0 & 173) & 0 & 37 & (0 & 330) & 0 & 54 & (0 & 044) & $-0$ & 72 & (0 & 019) \\
min($\sigma(g-r)$)$_{\mu_r\le21.2}$ & 0 & 64 & (0 & 000) & 0 & 66 & (0 & 052) & 0 & 37 & (0 & 190) & $-0$ & 44 & (0 & 199) \\
max($\sigma(g-r)$)$_{20.0\le\mu_r\le21.0}$ & 0 & 86 & (0 & 000) & 0 & 80 & (0 & 010) & 0 & 77 & (0 & 001) & 0 & 19 & (0 & 602) \\
\enddata
\tablecomments{ Pearson correlation coefficients (and p-values) with the WISE [4.6] $-$ [12] color, for all galaxies (All; 32), elliptical galaxies (E; 9), elliptical + disky elliptical galaxies (E + disky E; 14), and barred + unbarred spiral galaxies (Spirals; 13), based on the pCMDs after the standard masking procedure. }
\label{wisecc}
\end{deluxetable*}

Next, we test the capability of pCMD color dispersion parameters to classify galaxy morphology, in Figure~\ref{class1}. The combination of color dispersion divides early- and late-type galaxies as nicely as or even better than the color index combination does. This indicates that the complexity of stellar populations at given $\mu_r$ as well as their mean age and metallicity is an important feature discriminating between early- and late-type galaxies.
After examining various combinations, we found that the best combination to discriminate galaxy morphology is min$(\sigma(g-r))_{\mu_r\le21.2}$ and max$(\sigma(g-r))_{20.0\le\mu_r\le21.0}$.
In Figure~\ref{class1}(b) (zoomed-in in Figure~\ref{classin}(a)), the elliptical galaxies are well separated from the spiral galaxies, even from the bulge-dominated spiral galaxies unlike in Figure~\ref{stat1}. This seems to be mainly because max$(\sigma(g-r))_{20.0\le\mu_r\le21.0}$ represents how complex the stellar populations are at the region where the disk component starts to surpass the bulge component in a typical late-type galaxy. It consequently reflects the disk dominance in a galaxy, because stellar populations in a disk tend to be much more complex than those in a bulge. See the plots at the second row in Figure~\ref{procscheme} (and Appendix~\ref{app1}) for the spatial areas covered by the pixels with $20.0\le\mu_r\le21.0$ in each galaxy.

However, although the min$(\sigma(g-r))_{\mu_r\le21.2}$ versus max$(\sigma(g-r))_{20.0\le\mu_r\le21.0}$ plot seems to distinguish early-type (E, disky E and S0) galaxies from spiral galaxies almost perfectly, the edge-on disk galaxies are still mixed with the elliptical galaxies in this plot. 
To distinguish them, a parameter that is not from a pCMD is necessary: the axis ratio (b/a).
Figure~\ref{class1}(d) (zoomed-in in Figure~\ref{classin}(b)) shows that the edge-on disk galaxies are separated from the early-type galaxies in the b/a versus max$(\sigma(g-r))_{20.0\le\mu_r\le21.0}$ plot, due to their small b/a values.
In summary, at least in our sample, the early-type galaxies are well distinguished from the other galaxies in the parameter space of min$(\sigma(g-r))_{\mu_r\le21.2}$, max$(\sigma(g-r))_{20.0\le\mu_r\le21.0}$, and axis ratio.
The pCMDs from the alternative masking procedure also give consistent results (Appendix~\ref{app2}).

In Figure~\ref{btr}, we compare the pCMD color dispersion parameters with B/T, one of the most frequently used indicators of galaxy morphology. The performance of B/T for morphological classification is not bad: the elliptical and spiral galaxies are mostly separated by B/T. However, B/T fails to distinguish between elliptical galaxies and bulge-dominated spiral galaxies. Since max$(\sigma(g-r))_{20.0\le\mu_r\le21.0}$ even distinguishes between the elliptical galaxies and bulge-dominated galaxies, the performance of B/T appears to be poorer than the color dispersion.
Note that the structural decomposition may be improved if one makes more efforts, such as more careful setup of masks and initial guess, and more various component-functions for more realistic fitting \citep{pen10,kim16}. However, such a possibility of improvement in structural decomposition by various technical efforts inversely highlights the strength of the pCMD approach, which is simple and hardly depends on various technical conditions.

\begin{figure}[!t]
\centering
\plotone{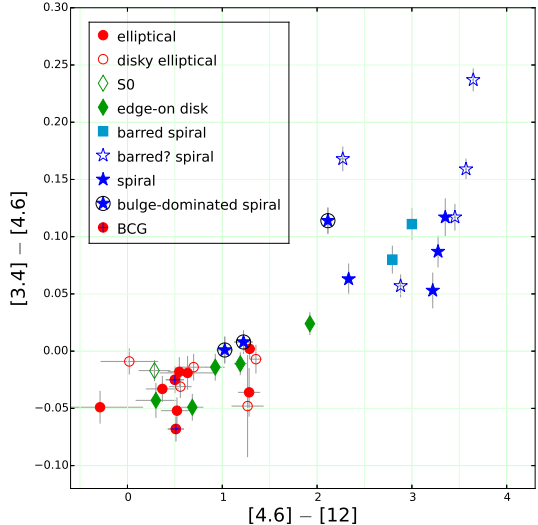}
\caption{ WISE infrared color-color diagram of the sample galaxies. Photometric uncertainties are denoted as gray lines for each target. \label{irccd}}
\end{figure}

\begin{figure*}
\centering
\includegraphics[width=0.95\textwidth]{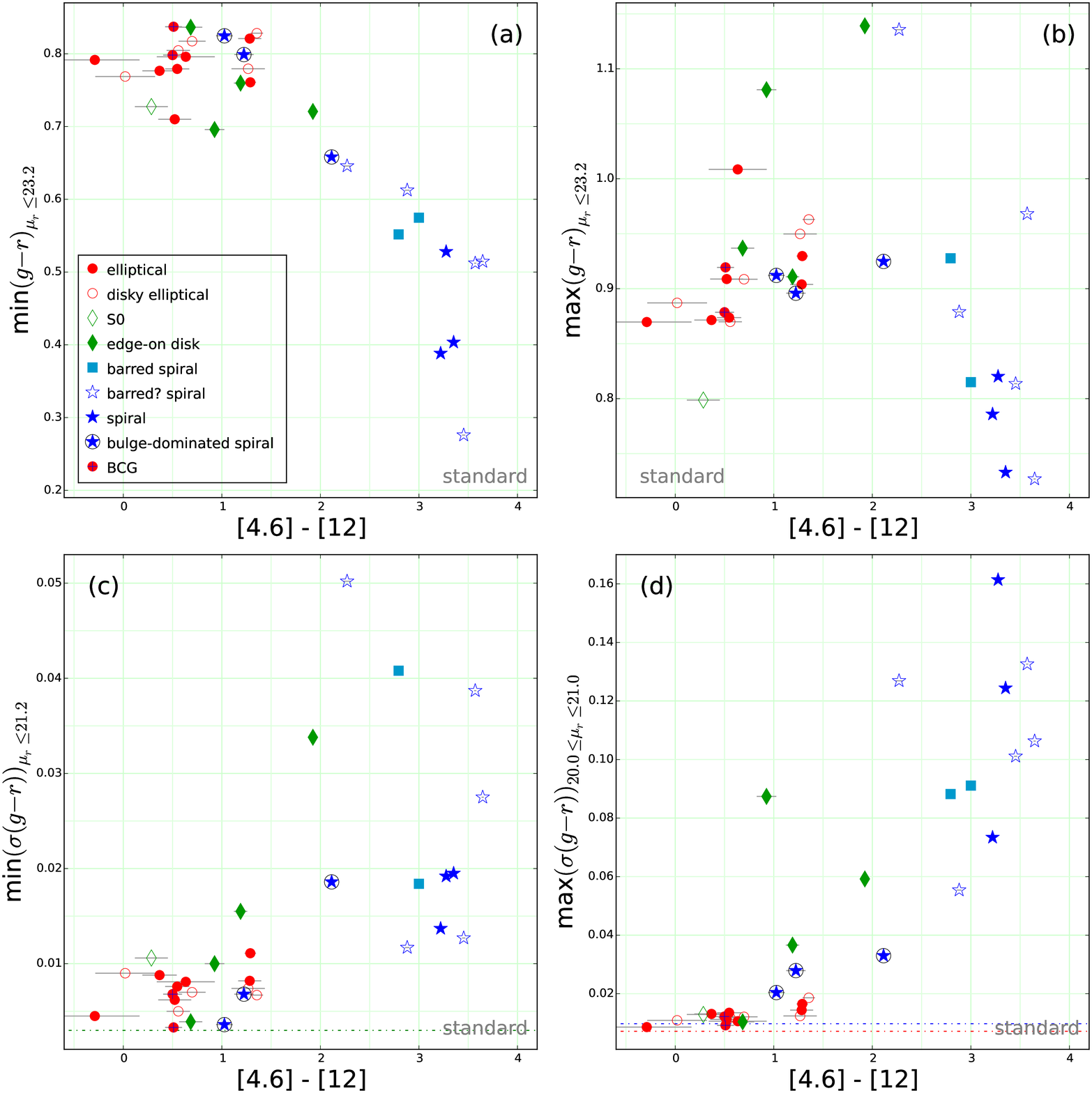}
\caption{ The pCMD parameters versus the WISE [4.6] $-$ [12] color, based on the pCMDs after the standard masking procedure: (a) the minimum $g-r$ color at $\mu_r\le23.2$ {\umu}, (b) the maximum $g-r$ color at $\mu_r\le23.2$ {\umu}, (c) the minimum $g-r$ color dispersion at $\mu_r\le21.2$ {\umu}, and (d) the maximum $g-r$ color dispersion at $20.0\le\mu_r\le21.0$ {\umu}. The dot-dashed lines are the reliability limits as described in Figure~\ref{coldev1}.\label{wise1}}
\end{figure*}

\subsection{Dependence on WISE Color}\label{wiseresult}

Another main purpose of this paper is to understand the relationship between stellar population complexity and recent star formation activity in a galaxy. While the stellar population complexity is measured by the pCMD color dispersion, we use photometric information in the infrared bands to estimate the star formation activities of the target galaxies.

\citet{jar17} showed how the infrared color-color ([3.4] $-$ [4.6] versus [4.6] $-$ [12]) diagram classifies galaxies according to their star formation and AGN activities.
Based on the scheme, Figure~\ref{irccd} shows that the morphological types of our sample galaxies are strongly correlated with their star formation activities. While all of our sample galaxies are in the non-AGN domain ([3.4] $-$ [4.6] $<$ 0.8), the spiral galaxies are clearly redder (more active star formation) than the elliptical galaxies in the [4.6] $-$ [12] color, except for two bulge-dominated spiral galaxies. Edge-on disk galaxies are in the intermediate domain between elliptical and spiral galaxies, although several elliptical galaxies and two bulge-dominated spiral galaxies also share that domain. Note that the edge-on disk galaxies in our sample tend to be bulge-dominated rather than disk-dominated, and thus they may be intrinsically similar to bulge-dominated spiral galaxies.

In Figure~\ref{wise1}, we compare the pCMD parameters and the total infrared colors of the sample galaxies.
There is a strong correlation between [4.6] $-$ [12] and min($g-r$)$_{\mu_r\le23.2}$, whereas max($g-r$)$_{\mu_r\le23.2}$ shows no clear correlation with [4.6] $-$ [12]. The former trend is as expected, because blue optical color and red infrared color are commonly the signals of star formation activity. Since star-forming spiral galaxies often have red bulges, the latter trend is also understood.
However, when we consider the elliptical galaxies only, no significant correlation is found between [4.6] $-$ [12] and min($g-r$)$_{\mu_r\le23.2}$. That is, the elliptical galaxies with small excess in infrared color show no difference in min($g-r$)$_{\mu_r\le23.2}$ from the elliptical galaxies without infrared-color excess.
On the other hand, for the elliptical galaxies, min($\sigma(g-r)$)$_{\mu_r\le21.2}$ and max($\sigma(g-r)$)$_{20.0\le\mu_r\le21.0}$ appear to be correlated with [4.6] $-$ [12]: the elliptical galaxies with larger [4.6] $-$ [12] tend to have larger min($\sigma(g-r)$)$_{\mu_r\le21.2}$ and max($\sigma(g-r)$)$_{20.0\le\mu_r\le21.0}$, as shown in Panels (c) and (d) of Figure~\ref{wise1}. 

The Pearson correlation coefficients and their p-values between several pCMD parameters and the WISE color are listed in Table~\ref{wisecc}. This table quantitatively shows that min($g-r$)$_{\mu_r\le23.2}$, min($\sigma(g-r)$)$_{\mu_r\le21.2}$ and max($\sigma(g-r)$)$_{20.0\le\mu_r\le21.0}$ are significantly correlated or anti-correlated with [4.6] $-$ [12], whereas max($g-r$)$_{\mu_r\le23.2}$ shows no significant correlation, for the whole sample. On the other hand, such trends are somewhat different when only elliptical galaxies are considered: min($\sigma(g-r)$)$_{\mu_r\le21.2}$ shows a marginal correlation and max($\sigma(g-r)$)$_{20.0\le\mu_r\le21.0}$ shows a more significant correlation.
It is also noted that max($g-r$)$_{\mu_r\le23.2}$ has a marginal correlation with [4.6] $-$ [12] when elliptical and disky elliptical galaxies are considered together, while it has an anti-correlation when spiral galaxies are considered only.
All of these trends do not significantly change even when we use the results from the alternative procedure (Appendix~\ref{app2}).

Table~\ref{fulltab1} lists the pCMD parameter values of the sample galaxies from the standard procedure and their WISE colors, which are used in Figures~\ref{coldev1} - \ref{wise1}. The WISE colors used in Appendix~\ref{app2} are the same as those in Table~\ref{fulltab1}.

\begin{turnpage}
\begin{deluxetable*}{lcccccccrr}
\tabletypesize{\scriptsize}
\tablenum{4} \tablecolumns{10} \tablecaption{pCMD Parameters from the Standard Procedure, and WISE Colors}\label{fulltab1} \tablewidth{0pt}
\tablehead{ Name & min$(g-r)$ & max$(g-r)$ & $\mu_r$(tip) & min$(\sigma(g-r))$ & max$(\sigma(g-r))$ & max$(\sigma(g-r))$ & max$(\sigma(g-r))$ & [3.4] $-$ [4.6] & [4.6] $-$ [12] \\
& $_{\mu_r\le23.2}$ & $_{\mu_r\le23.2}$ & & $_{\mu_r\le21.2}$ &  $_{20.0\le\mu_r\le21.0}$ & $_{18.5\le\mu_r\le19.5}$ & $_{\mu_r\le21.2}$ & & }
%\rotate
\startdata
A1139-00001 & 0.798 & 0.879 & 17.44 & 0.007 & 0.012 & 0.013 & 0.013 & $-0.025\pm0.012$ & $0.500\pm0.096$\\
A1139-00002 & 0.727 & 0.799 & 16.80 & 0.011 & 0.013 & 0.013 & 0.017 & $-0.017\pm0.009$ & $0.285\pm0.167$\\
A1139-00003 & 0.710 & 0.909 & 17.98 & 0.006 & 0.011 & 0.012 & 0.016 & $-0.052\pm0.012$ & $0.522\pm0.167$\\
A1139-00004 & 0.552 & 0.928 & 18.29 & 0.041 & 0.088 & 0.071 & 0.103 & $0.080\pm0.012$ & $2.794\pm0.026$\\
A1139-00005 & 0.805 & 0.870 & 17.35 & 0.005 & 0.012 & 0.012 & 0.012 & $-0.031\pm0.010$ & $0.558\pm0.118$\\
A1139-00006 & 0.613 & 0.879 & 17.77 & 0.012 & 0.055 & 0.037 & 0.055 & $0.057\pm0.010$ & $2.881\pm0.022$\\
A1139-00007 & 0.761 & 0.930 & 17.24 & 0.011 & 0.017 & 0.048 & 0.049 & $0.002\pm0.009$ & $1.289\pm0.057$\\
A1139-00008 & 0.799 & 0.896 & 17.78 & 0.007 & 0.028 & 0.017 & 0.028 & $0.008\pm0.010$ & $1.224\pm0.097$\\
A1139-00009 & 0.658 & 0.925 & 17.81 & 0.019 & 0.033 & 0.049 & 0.049 & $0.114\pm0.011$ & $2.114\pm0.032$\\
A1139-00010 & 0.777 & 0.872 & 17.03 & 0.009 & 0.013 & 0.014 & 0.014 & $-0.033\pm0.011$ & $0.368\pm0.174$\\
A1139-00011 & 0.721 & 1.139 & 18.41 & 0.034 & 0.059 & 0.136 & 0.136 & $0.024\pm0.010$ & $1.924\pm0.041$\\
A1139-00012 & 0.779 & 0.874 & 17.18 & 0.008 & 0.013 & 0.015 & 0.015 & $-0.018\pm0.013$ & $0.547\pm0.122$\\
A1139-00013 & 0.760 & 0.911 & 17.71 & 0.015 & 0.037 & 0.052 & 0.052 & $-0.011\pm0.010$ & $1.190\pm0.067$\\
A1139-00014 & 0.769 & 0.887 & 17.32 & 0.009 & 0.011 & 0.017 & 0.017 & $-0.009\pm0.011$ & $0.019\pm0.303$\\
A1139-00015 & 0.388 & 0.786 & 19.14 & 0.014 & 0.073 & 0.016 & 0.074 & $0.053\pm0.016$ & $3.220\pm0.021$\\
A1139-00016 & 0.575 & 0.815 & 17.51 & 0.018 & 0.091 & 0.033 & 0.091 & $0.111\pm0.014$ & $3.000\pm0.024$\\
A1139-00017 & 0.512 & 0.968 & 18.54 & 0.039 & 0.133 & 0.116 & 0.136 & $0.159\pm0.009$ & $3.571\pm0.011$\\
A2589-00001 & 0.837 & 0.919 & 17.52 & 0.003 & 0.009 & 0.005 & 0.016 & $-0.068\pm0.011$ & $0.510\pm0.085$\\
A2589-00002 & 0.696 & 1.081 & 16.93 & 0.010 & 0.087 & 0.070 & 0.087 & $-0.014\pm0.012$ & $0.926\pm0.100$\\
A2589-00003 & 0.514 & 0.727 & 17.16 & 0.028 & 0.106 & 0.055 & 0.123 & $0.237\pm0.010$ & $3.647\pm0.011$\\
A2589-00004 & 0.276 & 0.814 & 17.35 & 0.013 & 0.101 & 0.032 & 0.101 & $0.117\pm0.012$ & $3.454\pm0.016$\\
A2589-00005 & 0.404 & 0.733 & 18.20 & 0.019 & 0.124 & 0.044 & 0.124 & $0.117\pm0.017$ & $3.352\pm0.023$\\
A2589-00006 & 0.792 & 0.870 & 17.39 & 0.004 & 0.009 & 0.009 & 0.009 & $-0.049\pm0.014$ & $-0.289\pm0.454$\\
A2589-00007 & 0.825 & 0.912 & 17.54 & 0.004 & 0.020 & 0.010 & 0.020 & $0.001\pm0.012$ & $1.026\pm0.085$\\
A2589-00008 & 0.828 & 0.963 & 17.51 & 0.007 & 0.019 & 0.048 & 0.048 & $-0.007\pm0.013$ & $1.355\pm0.063$\\
A2589-00010 & 0.821 & 0.904 & 17.71 & 0.008 & 0.014 & 0.032 & 0.032 & $-0.036\pm0.021$ & $1.283\pm0.117$\\
A2589-00011 & 0.528 & 0.820 & 18.82 & 0.019 & 0.161 & 0.029 & 0.161 & $0.087\pm0.014$ & $3.276\pm0.024$\\
A2589-00012 & 0.817 & 0.909 & 17.58 & 0.007 & 0.012 & 0.010 & 0.015 & $-0.014\pm0.012$ & $0.698\pm0.137$\\
A2589-00013 & 0.837 & 0.937 & 17.02 & 0.004 & 0.010 & 0.006 & 0.011 & $-0.049\pm0.012$ & $0.684\pm0.117$\\
A2589-00014$^{*}$ & 0.779 & 0.950 & 17.66 & 0.007 & 0.012 & 0.023 & 0.023 & $-0.048\pm0.045$ & $1.267\pm0.169$\\
A2589-00015 & 0.646 & 1.136 & 18.46 & 0.050 & 0.127 & 0.157 & 0.157 & $0.168\pm0.011$ & $2.271\pm0.033$\\
A2589-00018 & 0.796 & 1.008 & 17.75 & 0.008 & 0.011 & 0.022 & 0.023 & $-0.019\pm0.015$ & $0.634\pm0.294$\\
\enddata
\tablecomments{ $^{*}$ The WISE colors of A2589-00014 is based on the profile-fit magnitudes, whereas those of all the other targets are based on the elliptical aperture magnitudes. See Section~\ref{wisedata} for details.}
\end{deluxetable*}
\end{turnpage}

\section{DISCUSSION}

\subsection{Issues on Masking}

There are some fundamental difficulties in masking contaminants.
In the case of early-type galaxies or edge-on disk galaxies, the masking is relatively easy: the SE-detection-based masking works well, because they hardly have confusing internal substructures. After the SE-detection-based masking, possibly remaining contaminants will be covered by the pCMD outlier masking, although it may sometimes over-mask internal substructures such as tidal debris.

What does matter is the late-type galaxies. 
Basically, it is very difficult to perfectly mask contaminants in and around a face-on late-type galaxy with complex spiral arms, because their substructures and contaminants are often too similar to be distinguished. Thus, in the standard procedure, we omitted the SE-detection-based masking for such targets, but only conducted the pCMD outlier masking that works alone to some extent.
However, this means that different masking processes are applied to early- and late-type galaxies, which may cause unfair comparison of the pCMD features between galaxies with different morphological types.

The alternative masking procedure was devised to complement such weakness of the standard procedure. Through the alternative procedure, fair comparison is possible between galaxies with any morphological types. However, the $R(\mu_r)$-based masking is less efficient than the SE-detection-based masking for early-type galaxies. It may fail to mask faint contaminants or contaminants very close to the target galaxy center.
Moreover, the $R(\mu_r)$-based masking may not cover the faint outskirt of a contaminating object due to the limit of the masking algorithm. Thus, it is recommendable to rely on the standard masking procedure if the sample consists of only early-type and edge-on galaxies. The main purpose of the alternative masking procedure is to check how reliable the results from the standard procedure are, by comparing them with the results from a self-consistent masking procedure.

Section~\ref{qcomp} and Appendix~\ref{app2} show that the results from the standard procedure and from the alternative procedure are not significantly different from each other. This is because our analysis in this paper focuses on the major trend in the pCMD of each galaxy rather than its fine features. The consistency between the results from the different procedures, despite the weakness of each masking procedure, indicates that the major trend of a pCMD is hardly influenced by the details of the masking methods.

\begin{figure*}[!t]
\centering
\includegraphics[width=0.95\textwidth]{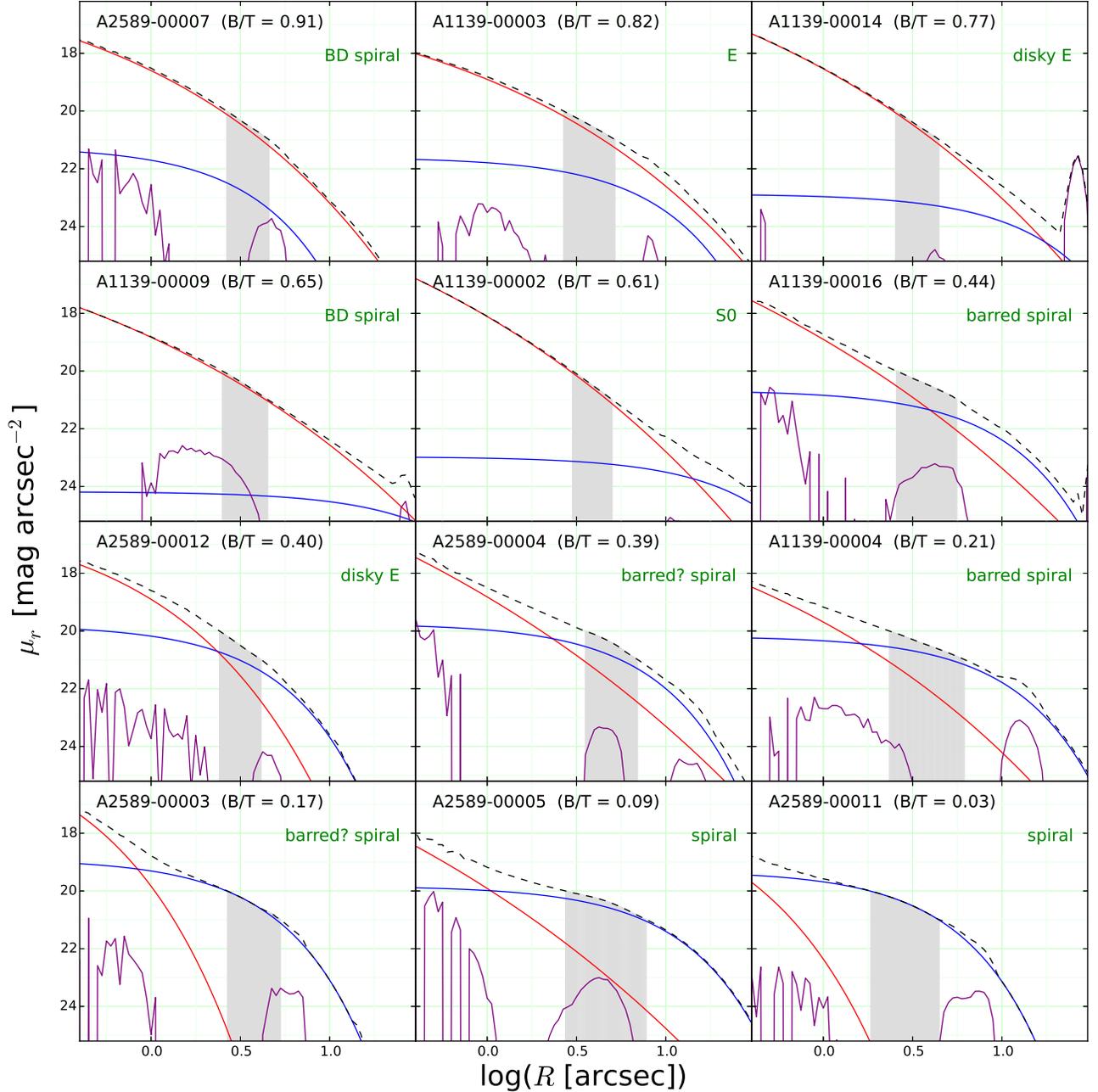}
\caption{ Radial profiles with structural decomposition using the GALFIT, for twelve galaxies having various B/T ratios ($0.03 - 0.91$). Each galaxy is decomposed using a Sersic component (bulge; red line) and a exponential component (disk; blue line). The purple line shows the residual light and the black dashed line is the total (bulge + disk + residual) profile in each panel. The radial regime corresponding to the total surface brightness between $20.0 - 21.0$ {\umu} is shaded on each plot.\label{profiles}}
\end{figure*}

\subsection{Morphological Segregation}

We tested various combinations of parameters measured from the pCMDs of our sample galaxies, to examine how galaxies with different morphological types are quantitatively distinguished in their pCMD features.
As a result, we found that the best parameter set to classify galaxy morphology based on pCMDs is the combination of the minimum color dispersion at $\mu_r\le21.2$ {\umu} and the maximum color dispersion at $20.0\le\mu_r\le21.0$ {\umu}, assisted by axis ratio (Figure~\ref{classin}).

Although these parameters were empirically selected, the underlying mechanism in which they work is relatively easy to understand.
As revealed in the pCMDs and the $\mu_r$ contour maps, max$(\sigma(g-r))_{20.0\le\mu_r\le21.0}$ represents how complex the stellar populations are at the regions where the disk component begins to be dominant in a typical late-type galaxy. In other words, max$(\sigma(g-r))_{20.0\le\mu_r\le21.0}$ is dominated by the bulge stellar populations for an early-type galaxy, whereas it is largely affected by the disk stellar populations for a late-type galaxy.
That makes max$(\sigma(g-r))_{20.0\le\mu_r\le21.0}$ a good indicator of disk dominance, because disks tend to have complex stellar populations spatially not uniform whereas bulges and elliptical galaxies have relatively uniform stellar populations.

Figure~\ref{profiles} shows twelve examples of radial profiles with structural decomposition using the GALFIT, which partially supports this interpretation (but not perfectly). For galaxies with B/T $>0.6$, the stellar populations at $20.0\le\mu_r\le21.0$ {\umu} are dominated by bulge components. On the other hand, for galaxies with B/T $<0.5$, the stellar populations at $20.0\le\mu_r\le21.0$ {\umu} are affected more significantly by disk components.

Nevertheless, the bulge-disk decomposition does not perfectly explain the performance of max$(\sigma(g-r))_{20.0\le\mu_r\le21.0}$, because there are some exceptional cases. For example, some bulge-dominated spiral galaxies (A1139-00009 and A2589-00007) have large B/T ratios and their stellar populations at $20.0\le\mu_r\le21.0$ {\umu} seem to be dominated by their bulge components. Despite the fact that their B/T ratios are even larger than that of the S0 galaxy A1139-00002 (0.65 and 0.91 versus 0.61), their max$(\sigma(g-r))_{20.0\le\mu_r\le21.0}$ values are larger than that of the S0 galaxy (0.036 and 0.020 versus 0.015). On the other hand, a disky elliptical galaxy A2589-00012 has a relatively small B/T (0.40), but its max$(\sigma(g-r))_{20.0\le\mu_r\le21.0}$ is as small as 0.015.
This indicates that the bulge-disk decomposition is not the only reason that max$(\sigma(g-r))_{20.0\le\mu_r\le21.0}$ works. A spiral galaxy, even though it is extremely bulge-dominated (e.g., A2589-00007), seems to have more complex stellar populations compared to an early-type galaxy with a similar B/T ratio. That is, the pCMD color dispersion is a good indicator to detect such fine difference in stellar populations that is hardly caught in classical methods of structural decomposition.

Note that $\mu_r=21.2$ {\umu} is the analysis limit for color dispersion in our data and thus the performance of color dispersion at $\mu_r>21.2$ {\umu} cannot be probed in this paper. If deeper images with sufficient S/N even at $\mu_r>21.2$ {\umu} are used, the usefulness of color dispersion as an indicator of galaxy morphology may be even larger, because the photometric properties of fainter part in a disk will be reflected. For example, in Figure~\ref{profiles}, the disk component of A1139-00002 (S0) is small, but it becomes relatively dominant at $R>10''$ (total $\mu_r\gtrsim22.2$ {\umu}). Thus, it may be useful even in distinguishing between elliptical and S0 galaxies if the color dispersion at lower surface brightness is available.

On the other hand,  min$(\sigma(g-r))_{\mu_r\le21.2}$ represents the minimum complexity of stellar populations, typically (but not necessarily) at the brightest center. Although elliptical galaxies and bulges of late-type galaxies typically have simple structures, they may show unusual substructures sometimes, which result in large color dispersion at their pCMD tips (e.g., A1139-00007 and A2589-00008). Since such substructures are thought to originate from mergers or interactions, min$(\sigma(g-r))_{\mu_r\le21.2}$ can be used as a simple indicator of recent mass assembly events of a target galaxy: larger min$(\sigma(g-r))_{\mu_r\le21.2}$ for more complex history of recent mass assembly on average.
This may not be a very precise indicator but will be useful particularly in a statistical study, because it can be conveniently applied to a bulk of galaxies, if appropriate corrections are conducted for the difference in redshift and spatial resolution.

It is interesting that the best parameter to characterize galaxy morphology is color dispersion rather than color index itself. In other words, for morphological classification of galaxies, it is more effective to compare how well mixed the stellar populations are rather than to see how old or metal-rich on average they are. The latter also works to some extent, but it fails to discriminate bulge-dominated spiral galaxies from elliptical galaxies in our results. This is not strange because the correlation between galaxy morphology and galaxy color is known to be not very tight \citep[e.g., blue early-type galaxies and red late-type galaxies;][]{lee06,lee08,toj13}.

However, for the generalized application of the morphological classification using the pCMD features, additional work is required. First of all, a much larger number of galaxies with well-classified morphological types need to be tested to get more reliable criteria of morphological classification. Our sample size of 32 is absolutely insufficient to establish criteria that can be generally applied.
Furthermore, it is necessary to examine how the criteria of morphological classification change as a function of image resolution. Since the pCMD features strongly depend on image resolution \citep{lee11,lee12,con16}, its effect on the pCMD classification should be seriously considered, to apply this method to galaxies at various redshifts.
Nevertheless, if these issues are addressed, the pCMD classification method will be a valuable technique to automatically classify a huge number of galaxies in the future imaging surveys using next generation facilities such as the Large Synoptic Survey Telescope \citep[LSST;][]{ive08}.

\subsection{Recent Growth of Elliptical Galaxies}

In Section~\ref{wiseresult}, the elliptical galaxies show a marginal correlation between their min($\sigma(g-r)$)$_{\mu_r\le21.2}$ and [4.6] $-$ [12] (Figure~\ref{wise1}(c)) and a stronger correlation between max($\sigma(g-r)$)$_{20.0\le\mu_r\le21.0}$ and [4.6] $-$ [12] (Figure~\ref{wise1}(d)).
Since the color dispersion reflects how complex stellar populations are in a galaxy, this result indicates that the elliptical galaxies with more complex stellar populations tend to have recently experienced more active star formation (or vice versa).
According to some previous studies, the recent growth of massive elliptical galaxies mainly depend on dry mergers rather than gas-rich mergers \citep{naa06,kor09,ber11}.
On the other hand, some other studies found the evidence of recent star formation activities probably triggered by mergers in some massive elliptical galaxies \citep{kav09,fer11,she16}.
Our result supports the latter findings: the recent star formation activity and mass growth of massive elliptical galaxies appear to be correlated with each other in our sample. Thus, the recent growth of those elliptical galaxies may not entirely depend on dry mergers, but gas-rich minor mergers may have usually happened.

One may suspect that the complexity of stellar populations in those galaxies may not necessarily originate from merger events, but it may be naturally caused by recent star formation not from merger origins. Even in that case, however, the recent star formation must have been spatially not uniform at scales larger than 600 pc ($\sim0.8''$ in our images) to enlarge the color dispersion in a pCMD. It may require anisotropic infall of gas. Gas interactions without disturbance of stellar orbits can be such an origin, which may happen even for cluster galaxies with high encounter velocities \citep{par09}. Although the exact origin of the recent star formation cannot be determined in this paper, it is plausible that some environmental effects acted on the large color dispersion in a pCMD.

The interpretation of the max($g-r$)$_{\mu_r\le23.2}$ versus [4.6] $-$ [12] plot (Figure~\ref{wise1}(b)) is somewhat complicated (Figure~\ref{wise1}(b)). At first glance, the data points seem to be randomly scattered in this plot.
However, as shown in Table~\ref{wisecc}, we find two opposite trends when we inspect the sub-samples of the elliptical + disky elliptical galaxies and the spiral galaxies: (1) the elliptical galaxies with infrared-color excess tend to have redder maximum pCMD color, but (2) on the other hand, the spiral galaxies are getting bluer in the optical band as infrared color increases.

These two opposite trends imply that multiple physical origins may be involved in the max($g-r$)$_{\mu_r\le23.2}$ versus [4.6] $-$ [12] relation.
For example, the marginal correlation of elliptical + disky elliptical galaxies may result from dust remnants in some elliptical galaxies. That is, elliptical galaxies that have recently experienced star formation may have remaining dust, which cause red infrared color (due to remaining warm dust) and red optical color (by dust extinction) at the same time.
On the other hand, spiral galaxies with redder infrared color may have more active current star formation unlike elliptical galaxies, and thus they tend to have blue optical color due to a large amount of young stars that cannot be sufficiently obscured by dust.
It will be worth checking if these trends are established still in a sufficiently large sample of galaxies in future studies.

\section{CONCLUSION}

We analyzed the pixel color-magnitude diagrams (pCMDs) of 32 bright galaxies in two galaxy clusters A1139 and A2589 at low redshifts.
We yielded the pCMDs of the sample galaxies using the SE-detection-based masking or the $R(\mu_r)$-based masking, any of which is complemented by the subsequent pCMD outlier masking process. We compared the results from the two procedures with each other, and confirmed that the different masking methods do not significantly affect the major trends of pCMD properties.

Our main conclusions from the pCMD analysis of the target galaxies are as follows:
\begin{enumerate}
 \item At least in our sample, the early- and late-type galaxies are most clearly separated by the combination of the minimum color dispersion at $\mu_r\le21.2$ {\umu} and the maximum color dispersion at $20.0\le\mu_r\le21.0$ {\umu}, while the edge-on galaxies are discriminated by using axis ratio. This is because min$(\sigma(g-r))_{\mu_r\le21.2}$ represents the minimum complexity of stellar populations typically at the brightest center, whereas max$(\sigma(g-r))_{20.0\le\mu_r\le21.0}$ reflects the complexity of stellar populations at the disk component in a typical spiral galaxy.
 \item The color dispersion measurements (min($\sigma(g-r)$)$_{\mu_r\le21.2}$ and max($\sigma(g-r)$)$_{20.0\le\mu_r\le21.0}$) of an elliptical galaxy appear to be correlated with its total infrared color ($[4.6] - [12]$). This indicates that the complexity of stellar populations in an elliptical galaxy is closely related with its recent star formation activity. From this observational evidence, we infer that gas-rich minor mergers or gas interactions may have usually happened during the recent growth of massive elliptical galaxies.
\end{enumerate}

In both conclusions, the color dispersion in a pCMD is a key quantity that appears to be closely related with the structure and formation history of a galaxy. Unlike the various pCMD features that are more complicated such as the backbone curvature and the spatial distribution of outlying pixels \citep{lee17}, the color dispersion can be simply measured and applied to a bulk of galaxies, to trace their recent formation histories. However, the definition and usage of this parameter need to be improved and more variously examined, because it must depend on rest-frame wavelength and spatial resolution of target images. Our follow-up studies using the full sample of the KYDISC clusters will cover this topic.

\acknowledgments

We appreciate the anonymous referee who gave us comments very helpful to improve this paper.
In this work, we used the data obtained under the K-GMT Science Program funded through Korea GMT Project operated by Korea Astronomy and Space Science Institute (KASI).
Parts of this research were conducted by the Australian Research Council Centre of Excellence for All Sky Astrophysics in 3 Dimensions (ASTRO 3D), through project number CE170100013.

\clearpage
\appendix
\section{A. $\,$ Full Plots in the Masking Processes}\label{app1}

\begin{figure*}
\centering
\includegraphics[width=0.95\textwidth]{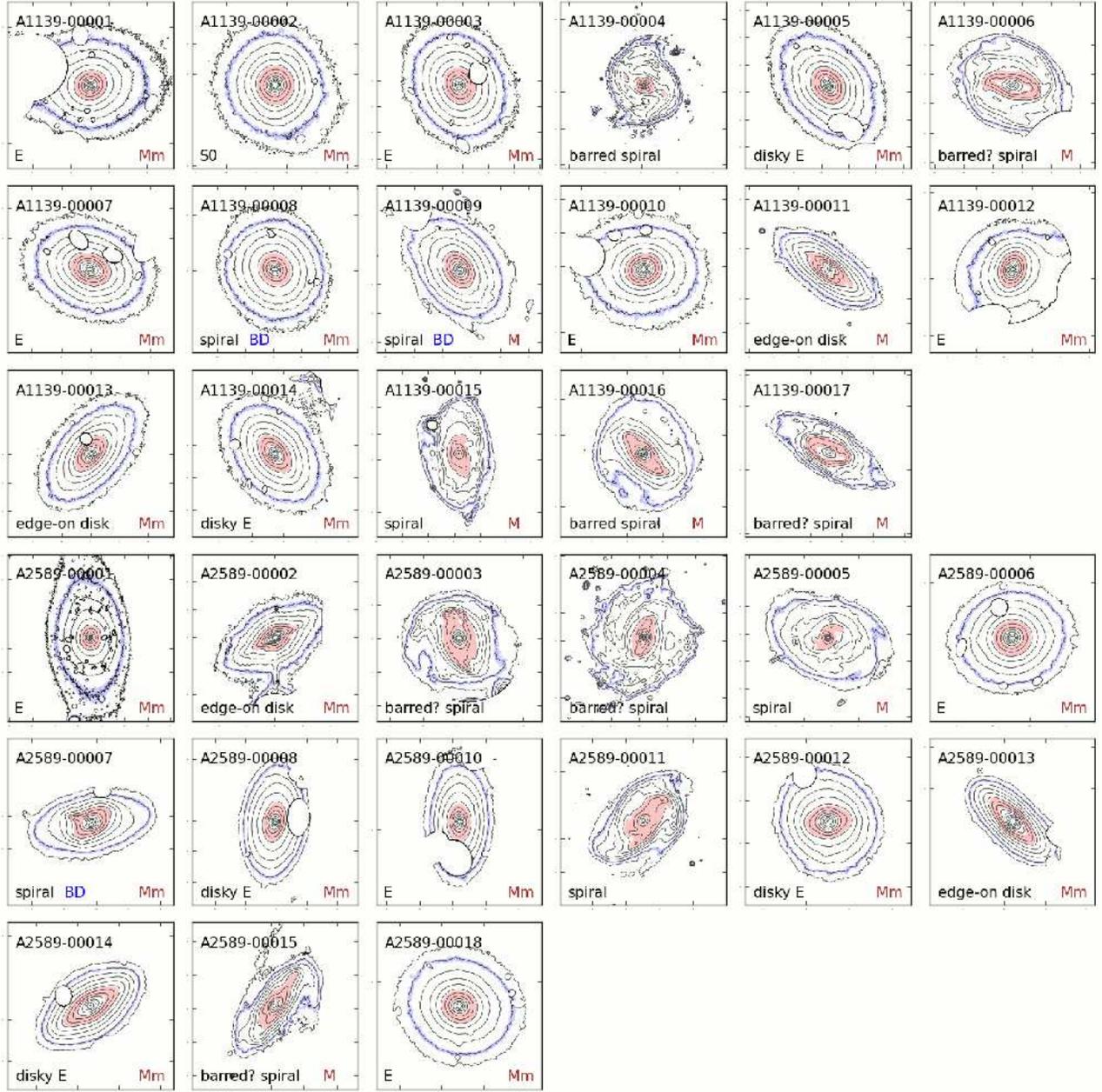}
\caption{Surface brightness contour maps after the SE-detection-based masking. The letters at the lower right corner indicates the sort of masking: `M' indicates that the target galaxy was masked using the SE detection with large background meshes only, while `Mm' indicates that the target galaxy was masked using the SE detection with small background meshes (m) as well as the SE detection with large background meshes (M). The regions colored with faint red show the pixels with $20.0\le\mu_r\le21.0$ {\umu}, while the regions colored with faint blue denote the pixels on the analysis limit ($\mu_r=23.2$ {\umu}). \label{mask1}}
\end{figure*}

\begin{figure*}
\centering
\includegraphics[width=0.95\textwidth]{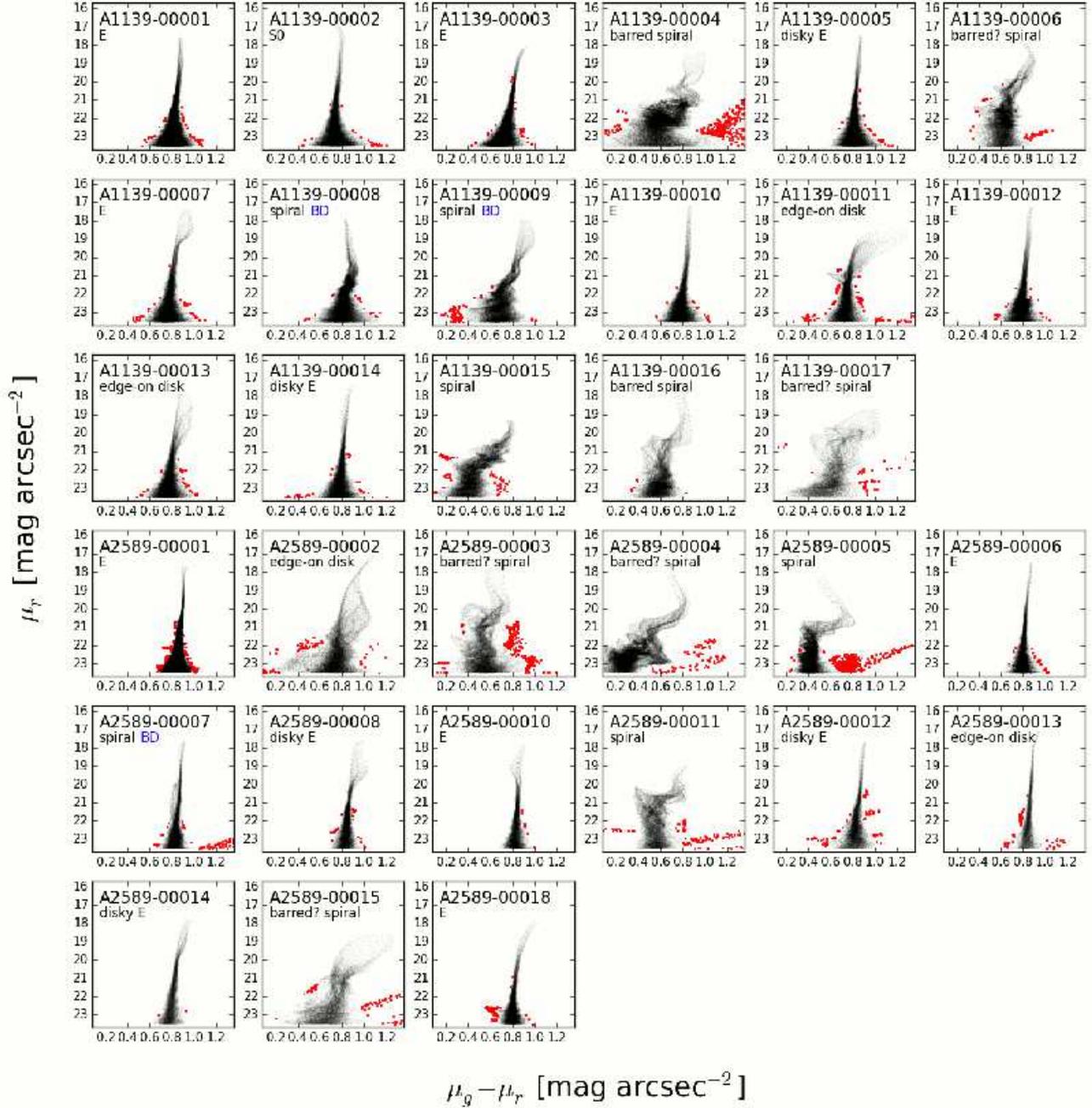}
\caption{The pCMDs after the SE-detection-based masking and the smoothing with the $0.8''$-aperture spline kernel. Red dots show the pCMD outliers (see Section~\ref{standard} for details).\label{pcmdout1}}
\end{figure*}

\begin{figure*}
\centering
\includegraphics[width=0.95\textwidth]{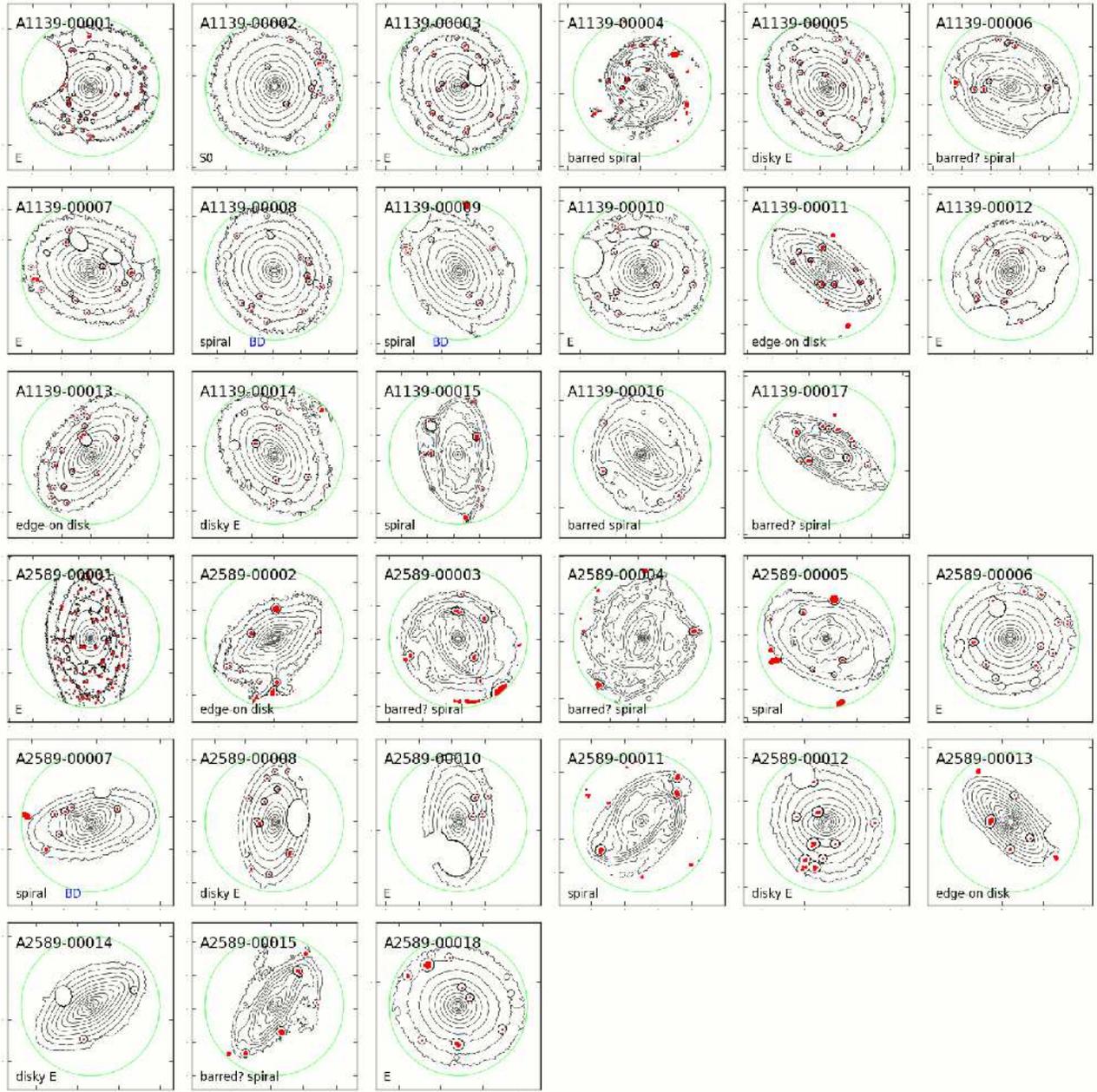}
\caption{Surface brightness contour maps after masking pCMD outliers and their neighboring pixels within $0.8''$. The outlying pixels defined in Figure~\ref{pcmdout1} are marked as red dots. The green circle indicates the spatial extent for the final pixel analysis of each target, which is defined at Step~2 of the alternative procedure in Section~\ref{altproc}.\label{finmask1}}
\end{figure*}

\begin{figure*}
\centering
\includegraphics[width=0.95\textwidth]{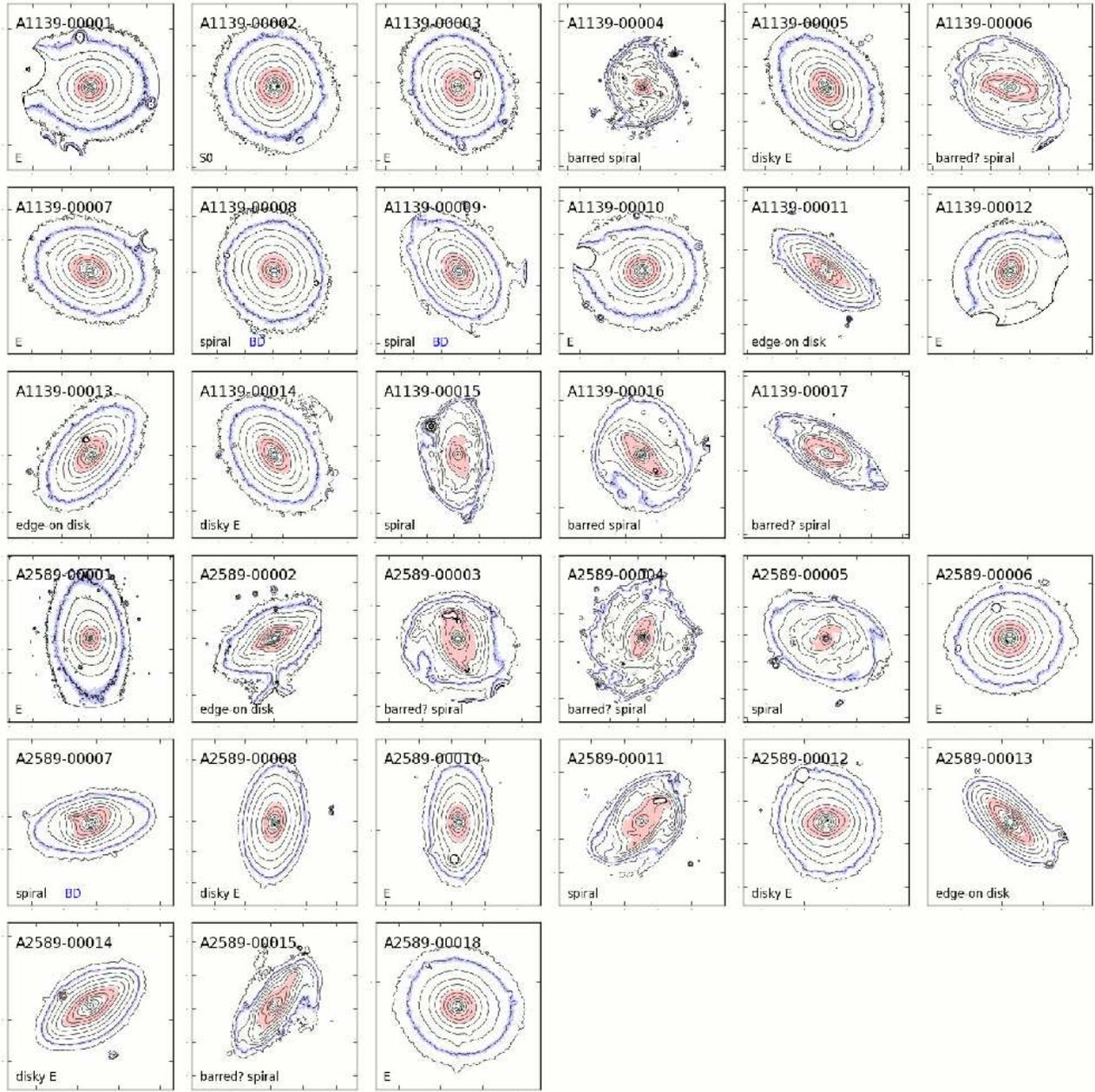}
\caption{Surface brightness contour maps after the $R(\mu_r)$-based masking: the same as Figure~\ref{mask1}, except that these are from the alternative masking procedure.\label{mask2}}
\end{figure*}

\begin{figure*}
\centering
\includegraphics[width=0.95\textwidth]{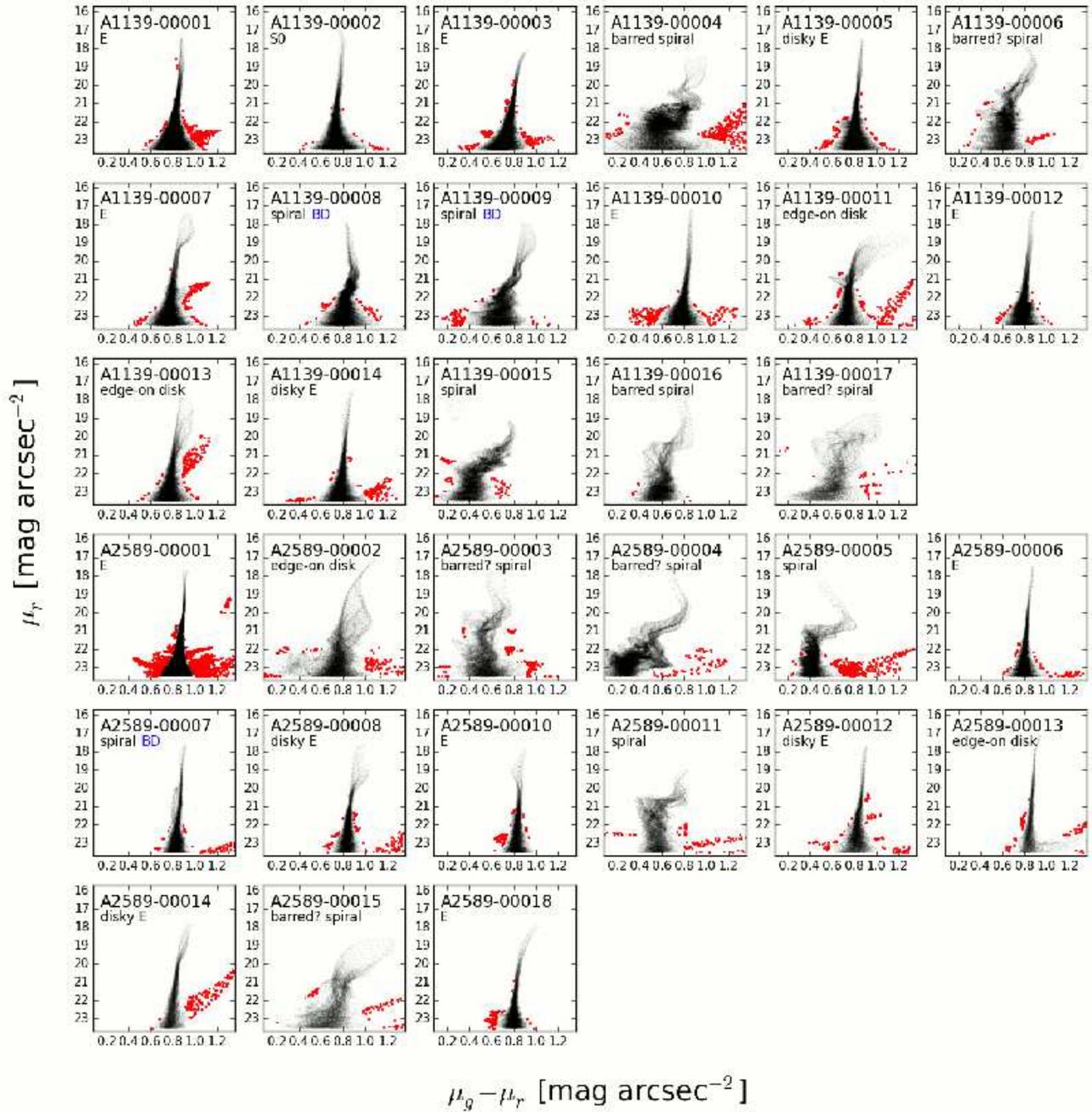}
\caption{The pCMDs after the $R(\mu_r)$-based masking and the smoothing with the $0.8''$-aperture spline kernel: the same as Figure~\ref{pcmdout1}, except that these are from the alternative masking procedure.\label{pcmdout2}}
\end{figure*}

\begin{figure*}
\centering
\includegraphics[width=0.95\textwidth]{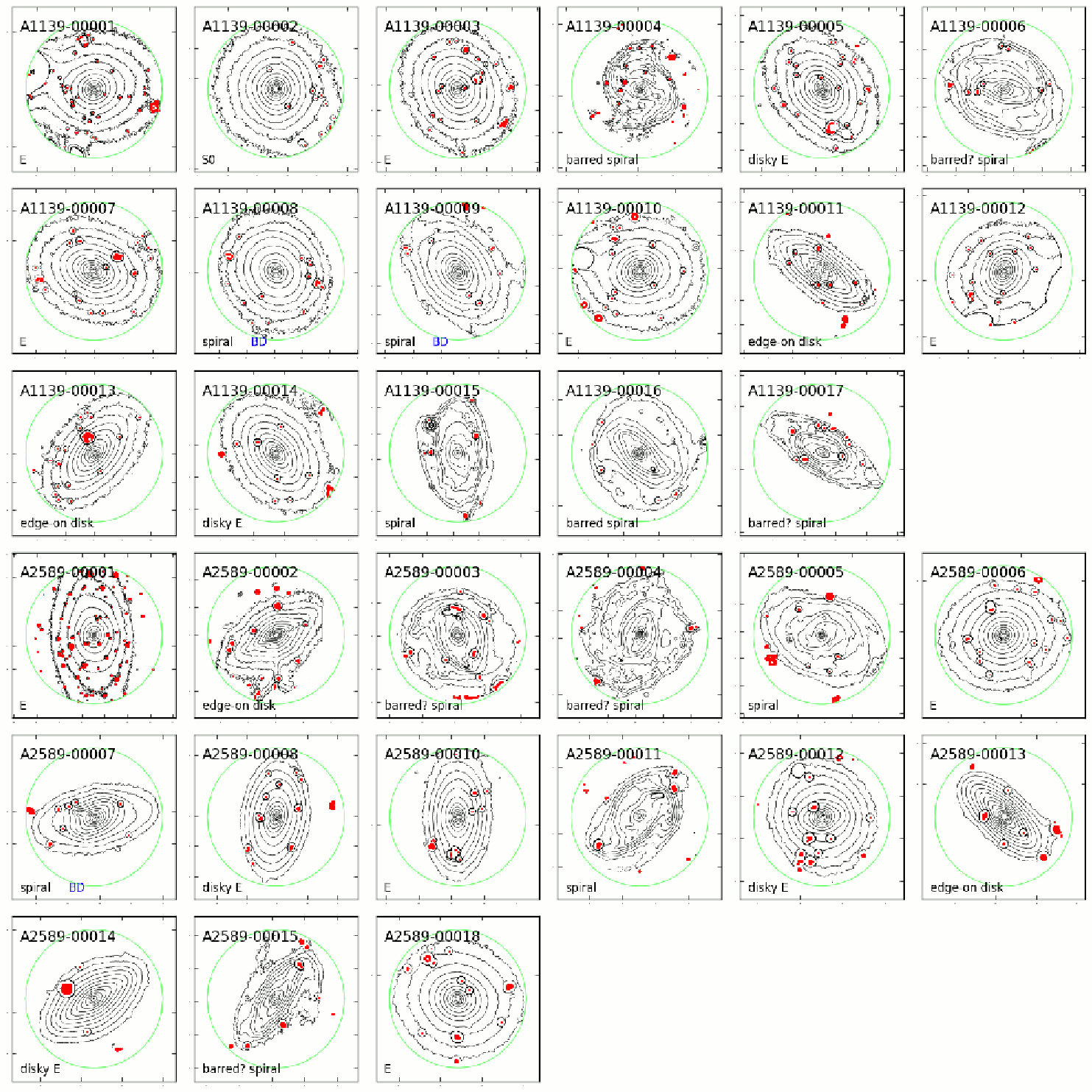}
\caption{Surface brightness contour maps after masking pCMD outliers and their neighboring pixels within $0.8''$: the same as Figure~\ref{finmask1}, except that these are from the alternative masking procedure.\label{finmask2}}
\end{figure*}

In Figure~\ref{procscheme}, we schematically summarized the masking processes in the standard and alternative procedures. However, the full plots showing the detailed masking processes for the whole sample galaxies were omitted to improve the readability of this paper. In this section, those full plots are presented.

The masking processes in the standard procedure flow as shown in Figures~\ref{mask1} -- \ref{finmask1}:
\begin{enumerate}
 \item The target galaxies are masked using the SE detection (Figure~\ref{mask1}). The SE-detection-based masking is adaptively applied with consideration of the morphology of a target galaxy and the performance of the masks. Among the 32 sample galaxies, 20 galaxies are masked using the SE detection with both of large and small background meshes (`Mm'), whereas eight galaxies are masked using the SE detection only with large background meshes (`M'). The rest four galaxies are not masked at all.
 \item Using the pCMDs after the SE-detection-based masking, outlying pixels are determined with the method described in Section~\ref{standard} (Figure~\ref{pcmdout1}).  Some of the selected outliers may be substructures of target galaxies, but the possible over-masking does not significantly change the major trends of pCMDs.
 \item The pCMD outliers and their neighboring pixels within $0.8''$ distance (seeing size) are additionally masked, because those neighboring pixels may also be affected by the contaminants (Figure~\ref{finmask1}). This additional masking is particularly important for late-type galaxies, which are poorly masked in the SE-detection-based method.
\end{enumerate}
The final pCMDs after these processes are shown in Figure~\ref{pcmdcon1}.

The full plots for the alternative procedure are also presented in Figures~\ref{mask2} -- \ref{finmask2}, and the final pCMDs are shown in Figure~\ref{pcmdcon2}.
Although the pCMDs resulting from the first-step masking in the two procedures is pretty different from each other (Figures~\ref{pcmdout1} and \ref{pcmdout2}),  the difference of the final pCMDs is smaller (Figures~\ref{pcmdcon1} -- \ref{dpcmd}), owing to the secondary masking process based on pCMD outliers.

\clearpage

\section{B. $\,$ Results from the pCMDs after the Alternative Masking Procedure}\label{app2}

\begin{figure*}[p]
\centering
\includegraphics[width=0.95\textwidth]{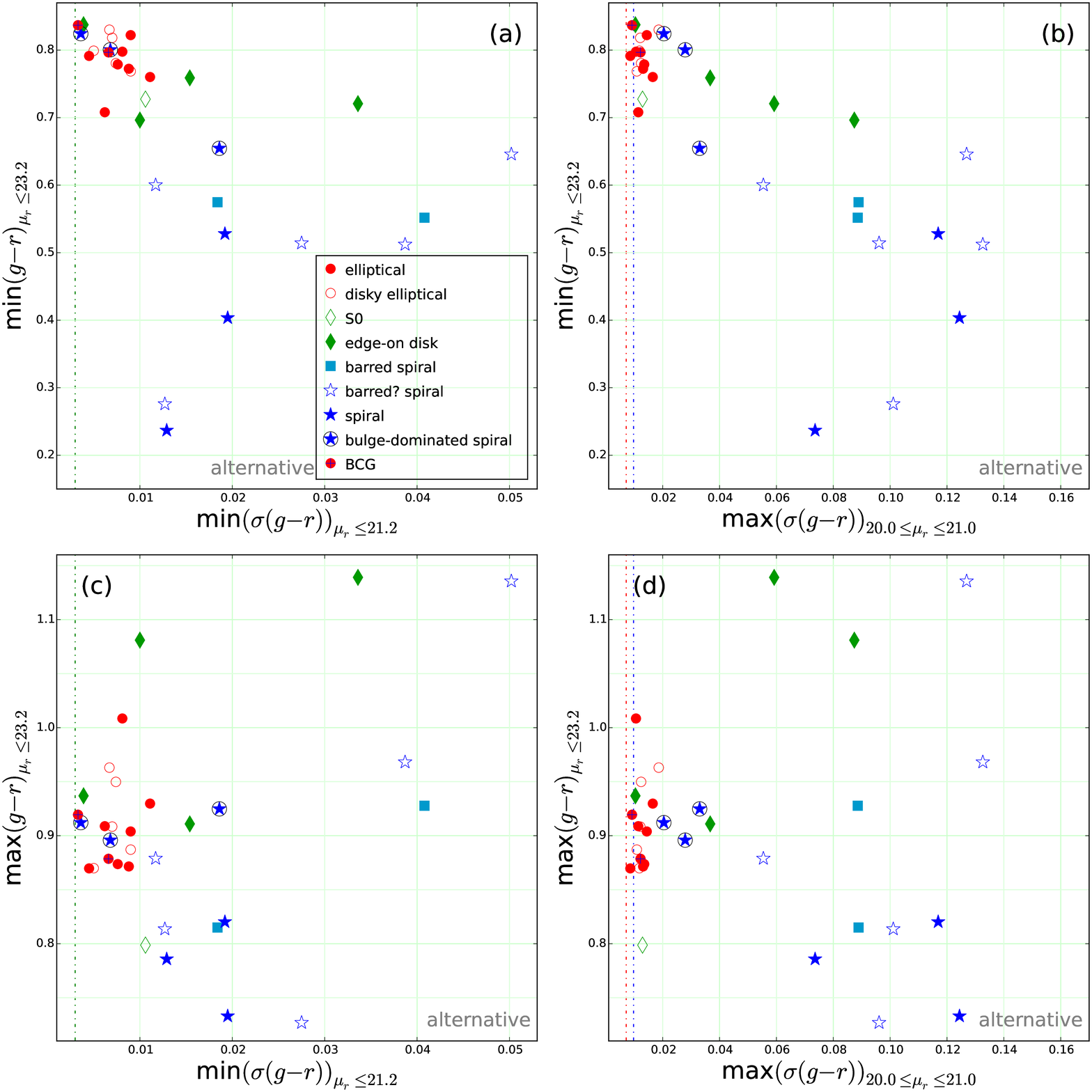}
\caption{Correlations between pCMD parameters, based on the pCMDs after the alternative procedure:
(a) the minimum $g-r$ color at $\mu_r\le23.2$ {\umu} versus the minimum $g-r$ color dispersion at $\mu_r\le21.2$ {\umu},
(b) the minimum $g-r$ color at $\mu_r\le23.2$ {\umu} versus the maximum $g-r$ color dispersion at $20.0\le\mu_r\le21.0$ {\umu},
(c) the maximum $g-r$ color at $\mu_r\le23.2$ {\umu} versus the minimum $g-r$ color dispersion at $\mu_r\le21.2$ {\umu}, and
(d) the maximum $g-r$ color at $\mu_r\le23.2$ {\umu} versus the maximum $g-r$ color dispersion at $20.0\le\mu_r\le21.0$ {\umu}. The symbols are the same as those in Figure~\ref{coldev1}.\label{coldev2}}
\end{figure*}

\begin{figure*}[!ht]
\centering
\includegraphics[width=0.95\textwidth]{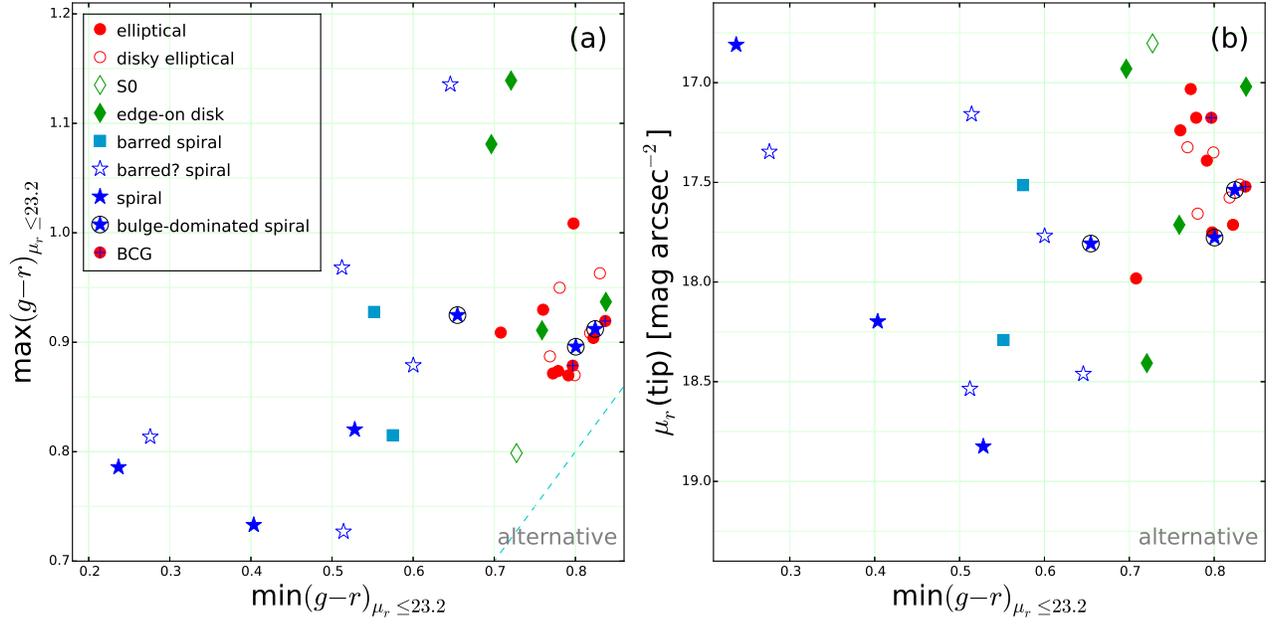}
\caption{(a) The maximum versus minimum values of mean $g-r$ color at $\mu_r\le23.2$ {\umu}, and (b) the brightest $\mu_r$ versus the minimum $g-r$ color at $\mu_r\le23.2$ {\umu}, based on the pCMDs after the alternative masking procedure. The symbols are the same as those in Figure~\ref{stat1}. \label{stat2}}
\end{figure*}

\begin{figure*}
\centering
\includegraphics[width=0.95\textwidth]{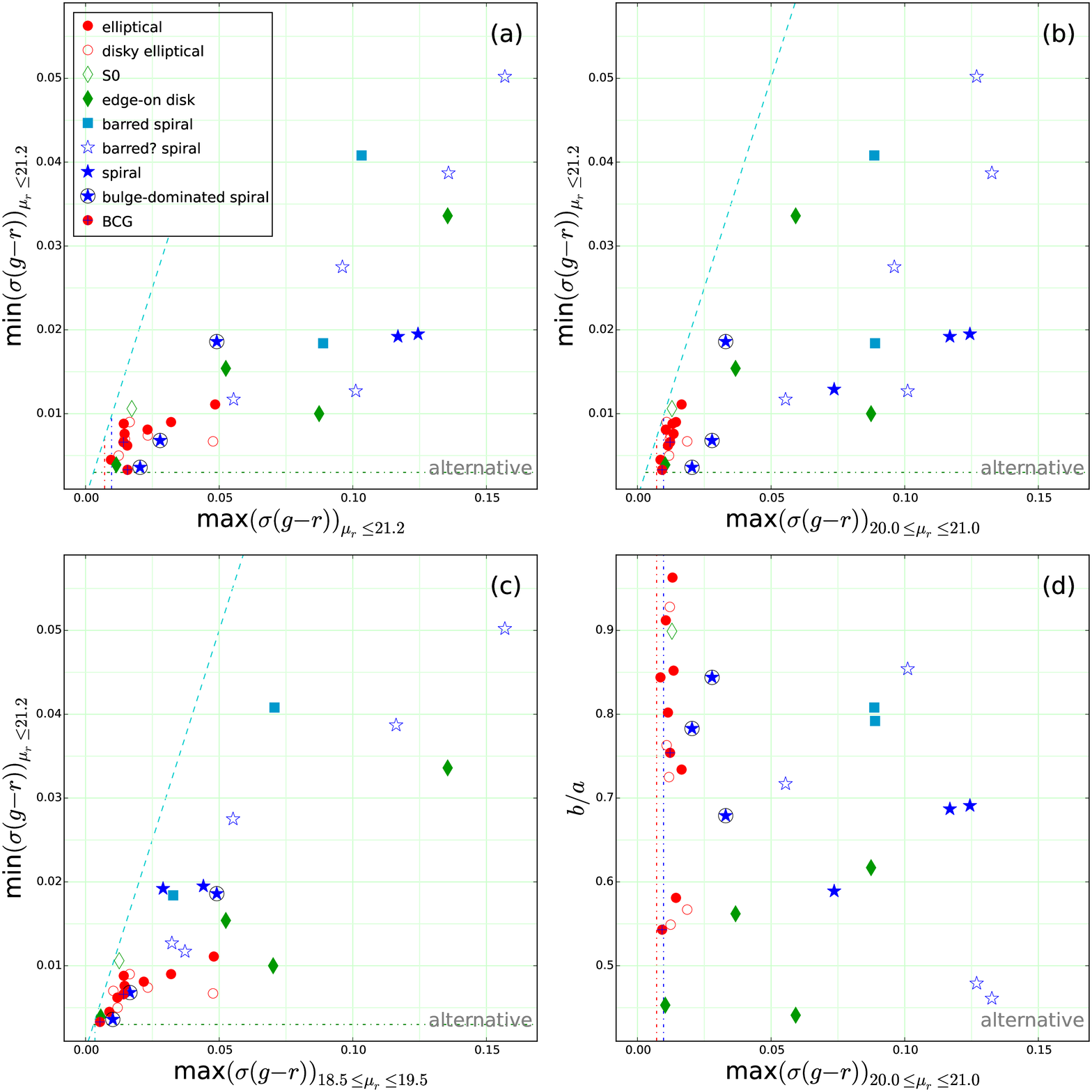}
\caption{The minimum $g-r$ color dispersion at $\mu_r\le21.2$ {\umu} (a) versus the maximum $g-r$ color dispersion at $\mu_r\le21.2$ {\umu}, (b) versus the maximum $g-r$ color dispersion at $20.0\le\mu_r\le21.0$ {\umu}, and (c) versus the maximum $g-r$ color dispersion at $18.5\le\mu_r\le19.5$ {\umu}. (d) Axis ratio versus the maximum $g-r$ color dispersion at $20.0\le\mu_r\le21.0$ {\umu}. These plots are the same as those in Figure~\ref{class1}, except that these are from the alternative masking procedure. \label{class2}}
\end{figure*}

\begin{figure*}
\centering
\includegraphics[width=0.95\textwidth]{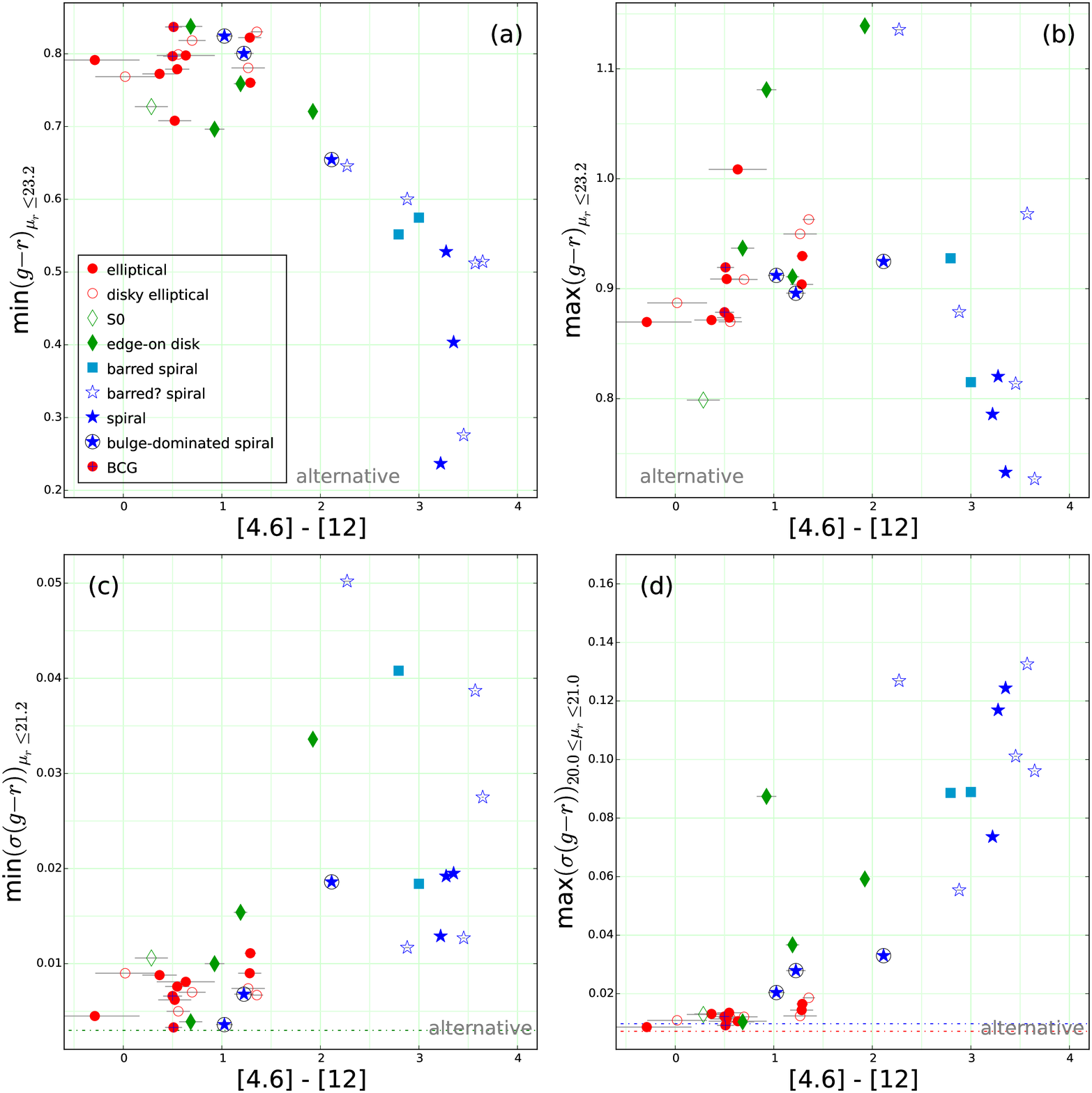}
\caption{The pCMD parameters versus the WISE [4.6] $-$ [12] color, based on the pCMDs after the alternative masking procedure: (a) the minimum $g-r$ color at $\mu_r\le23.2$ {\umu}, (b) the maximum $g-r$ color at $\mu_r\le23.2$ {\umu}, (c) the minimum value of $g-r$ color dispersion at $\mu_r\le21.2$ {\umu}, and (d) the maximum value of $g-r$ color dispersion at $20.0\le\mu_r\le21.0$ {\umu}. The dot-dashed lines are the reliability limits as described in Figure~\ref{coldev1}.\label{wise2}}
\end{figure*}

All figures in Section~\ref{result} are based on the pCMDs after the standard masking procedure, and the results from the alternative procedure were omitted for better readability. The differences between the final results from the two different procedures are not significant. In this section, those omitted results are presented.

Figures~\ref{coldev2}, \ref{stat2}, \ref{class2} and \ref{wise2} are the alternative procedure versions of Figures~\ref{coldev1}, \ref{stat1}, \ref{class1} and \ref{wise1}. The results from the alternative procedure show little difference from the results from the standard procedure.
Most sample galaxies show little changes in their distributions on the plots, particularly for elliptical galaxies.
Only a few spiral galaxies show considerable changes in their pCMD parameters (e.g., A1139-00015 and A2589-00011), which result from incomplete masking of contaminants that happens in spiral galaxies more easily.
However, those changes do not affect our main conclusions at all.
Table~\ref{fulltab2} lists the pCMD parameter values of the target galaxies from the alternative procedure, which are used in Figures~\ref{coldev2} - \ref{wise2}.

\begin{deluxetable*}{ccccccccrr}
\tabletypesize{\footnotesize}
\tablenum{5} \tablecolumns{8} \tablecaption{pCMD Parameters from the Alternative Procedure}\label{fulltab2} \tablewidth{0pt}
\tablehead{ Name & min$(g-r)$ & max$(g-r)$ & $\mu_r$(tip) & min$(\sigma(g-r))$ & max$(\sigma(g-r))$ & max$(\sigma(g-r))$ & max$(\sigma(g-r))$ \\
& $_{\mu_r\le23.2}$ & $_{\mu_r\le23.2}$ & & $_{\mu_r\le21.2}$ &  $_{20.0\le\mu_r\le21.0}$ & $_{18.5\le\mu_r\le19.5}$ & $_{\mu_r\le21.2}$ }
%\rotate
\startdata
A1139-00001 & 0.797 & 0.879 & 17.18 & 0.007 & 0.012 & 0.014 & 0.014\\ 
A1139-00002 & 0.727 & 0.799 & 16.80 & 0.011 & 0.013 & 0.013 & 0.017\\ 
A1139-00003 & 0.708 & 0.909 & 17.98 & 0.006 & 0.011 & 0.012 & 0.016\\ 
A1139-00004 & 0.552 & 0.928 & 18.29 & 0.041 & 0.089 & 0.071 & 0.103\\ 
A1139-00005 & 0.799 & 0.870 & 17.35 & 0.005 & 0.012 & 0.012 & 0.012\\ 
A1139-00006 & 0.600 & 0.879 & 17.77 & 0.012 & 0.055 & 0.037 & 0.055\\ 
A1139-00007 & 0.760 & 0.930 & 17.24 & 0.011 & 0.017 & 0.048 & 0.049\\ 
A1139-00008 & 0.800 & 0.896 & 17.78 & 0.007 & 0.028 & 0.017 & 0.028\\ 
A1139-00009 & 0.655 & 0.925 & 17.81 & 0.019 & 0.033 & 0.049 & 0.049\\ 
A1139-00010 & 0.772 & 0.872 & 17.03 & 0.009 & 0.013 & 0.014 & 0.014\\ 
A1139-00011 & 0.721 & 1.139 & 18.41 & 0.034 & 0.059 & 0.136 & 0.136\\ 
A1139-00012 & 0.779 & 0.874 & 17.18 & 0.008 & 0.013 & 0.015 & 0.015\\ 
A1139-00013 & 0.759 & 0.911 & 17.71 & 0.015 & 0.037 & 0.052 & 0.052\\ 
A1139-00014 & 0.769 & 0.887 & 17.32 & 0.009 & 0.011 & 0.017 & 0.017\\ 
A1139-00015 & 0.237 & 0.786 & 16.81 & 0.013 & 0.074 & 0.245 & 0.245\\ 
A1139-00016 & 0.575 & 0.815 & 17.51 & 0.018 & 0.089 & 0.033 & 0.089\\ 
A1139-00017 & 0.512 & 0.968 & 18.54 & 0.039 & 0.133 & 0.116 & 0.136\\ 
A2589-00001 & 0.837 & 0.919 & 17.52 & 0.003 & 0.009 & 0.005 & 0.016\\ 
A2589-00002 & 0.696 & 1.081 & 16.93 & 0.010 & 0.087 & 0.070 & 0.087\\ 
A2589-00003 & 0.514 & 0.727 & 17.16 & 0.028 & 0.096 & 0.055 & 0.096\\ 
A2589-00004 & 0.276 & 0.814 & 17.35 & 0.013 & 0.101 & 0.032 & 0.101\\ 
A2589-00005 & 0.404 & 0.733 & 18.20 & 0.019 & 0.124 & 0.044 & 0.124\\ 
A2589-00006 & 0.791 & 0.870 & 17.39 & 0.004 & 0.009 & 0.009 & 0.009\\ 
A2589-00007 & 0.824 & 0.912 & 17.54 & 0.004 & 0.020 & 0.010 & 0.020\\ 
A2589-00008 & 0.830 & 0.963 & 17.51 & 0.007 & 0.019 & 0.048 & 0.048\\ 
A2589-00010 & 0.822 & 0.904 & 17.71 & 0.009 & 0.014 & 0.032 & 0.032\\ 
A2589-00011 & 0.528 & 0.820 & 18.82 & 0.019 & 0.117 & 0.029 & 0.117\\ 
A2589-00012 & 0.818 & 0.908 & 17.58 & 0.007 & 0.012 & 0.010 & 0.015\\ 
A2589-00013 & 0.838 & 0.937 & 17.02 & 0.004 & 0.010 & 0.006 & 0.011\\ 
A2589-00014 & 0.781 & 0.950 & 17.66 & 0.007 & 0.012 & 0.023 & 0.023\\ 
A2589-00015 & 0.646 & 1.136 & 18.46 & 0.050 & 0.127 & 0.157 & 0.157\\ 
A2589-00018 & 0.798 & 1.008 & 17.75 & 0.008 & 0.011 & 0.022 & 0.023\\ 
\enddata
\end{deluxetable*}

\end{document}